\documentclass[preprintnumbers,superscriptaddress,nofootinbib,aps,prd,floatfix]{revtex4}

\usepackage{mathtools}
\usepackage{footmisc}
\usepackage{amsmath}
\usepackage{amsfonts}
\usepackage{amssymb}
\usepackage{wasysym}
\usepackage{graphicx}
\usepackage{color,array,subfigure}
\usepackage{comment}
\usepackage{latexsym}

\newcommand{\be}{\begin{eqnarray}}
\newcommand{\ee}{\end{eqnarray}}

\newcommand{\bee}{\begin{eqnarray}}
\newcommand{\eee}{\end{eqnarray}}
\newcommand{\beeq}{\begin{equation}}
\newcommand{\eeeq}{\end{equation}}

\renewcommand{\vec}{\bf}

\newcommand{\hfpi}{\hat{f}_\pi}

\numberwithin{equation}{section}

\begin{document}

\title{Sphalerons in composite and non-standard Higgs models}

\begin{abstract}
After the discovery of the Higgs boson and the rather precise measurement of all electroweak boson's masses the local structure of the electroweak symmetry breaking potential is already quite well established. However, despite being a key ingredient to a fundamental understanding of the underlying mechanism of electroweak symmetry breaking, the global structure of the electroweak potential remains entirely unknown.
The existence of sphalerons, unstable solutions of the classical action of motion that are interpolating between topologically distinct vacua, is a direct consequence of the Standard Model's $\mathrm{SU}(2)_L$ gauge group. Nevertheless, the sphaleron energy depends on the shape of the Higgs potential away from the minimum and can therefore be a litmus test for its global structure. Focusing on two scenarios, the minimal composite Higgs model $\mathrm{SO}(5)/\mathrm{SO}(4)$ or an elementary Higgs with a deformed electroweak potential, we calculate the change of the sphaleron energy compared to the Standard Model prediction. We find that the sphaleron energy would have to be measured to $\mathcal{O}(10)\%$ accuracy to exclude sizeable global deviations from the Standard Model Higgs potential. We further find that because of the periodicity of the scalar potential in composite Higgs models a second sphaleron branch with larger energy arises.

\end{abstract}

\author{Michael Spannowsky} \email{michael.spannowsky@durham.ac.uk}
\affiliation{Institute for Particle Physics Phenomenology, Department
  of Physics,\\Durham University, Durham DH1 3LE, UK\\[0.1cm]}

\author{Carlos Tamarit} \email{carlos.tamarit@durham.ac.uk}
\affiliation{Institute for Particle Physics Phenomenology, Department
  of Physics,\\Durham University, Durham DH1 3LE, UK\\[0.1cm]}

\preprint{IPPP/16/106}

\maketitle

\section{Introduction}
\label{sec:intro}
The recent discovery of the Higgs boson \cite{Aad:2012tfa, Chatrchyan:2012xdj} and the ongoing measurements of its properties \cite{Khachatryan:2016vau} are in good agreement with the hypothesis that this particle is a remnant of the Brout-Englert-Higgs mechanism, i.e. the spontaneous breaking of $\mathrm{SU}(2)_L \times \mathrm{U}(1)_Y \to \mathrm{U}(1)_\mathrm{QED}$. 

While the precise determination of the Higgs and gauge boson masses, as well as the interactions of the Higgs boson with elementary particles, including itself, will continue to improve our understanding of the scalar potential's local structure in the vicinity of the vacuum, its global structure, which can possibly explain the nature of electroweak symmetry breaking, is very difficult to probe experimentally. 

For example, the nature of the Higgs, whether elementary or composite, is still an open question. Even if the Higgs is assumed to be elementary, the shape of its potential remains unknown. It could be of mexican-hat shape as in 
the Standard Model (SM), or it could be deformed by strong quantum corrections due to virtual effects of additional fields. Were the Higgs boson to be a composite pseudo-Nambu-Goldstone boson of a strongly-coupled sector, one would expect
a periodic potential involving trigonometric functions. In all cases, the Higgs mass is fixed by the curvature of the potential at its minimum, and so in the vicinity of the latter the shape of the potential will be similar in all possible models. Nevertheless, deviations are allowed away from the minimum. For example, one could have a barrier at zero temperature between the vacuum and the origin of field-space. Moreover, in 
composite Higgs models the relation between the Higgs field's vacuum expectation value (VEV) and the gauge boson masses differs from its SM counterpart, and thus the location of the minimum in field-space may vary.

Discriminating between the different possibilities is of fundamental importance for our understanding of nature and, hence, the embedding of the effective Standard Model in an underlying UV theory. This motivates to consider possible observables which could be sensitive to the Higgs potential beyond its minimum. A possible candidate is the energy scale of baryon-number-violating processes. If baryon number is only violated by the anomaly under the weak interactions, then it follows that processes that violate baryon-number are associated with transitions between vacua classified by their weak topological charge. The minimum energy barrier between these vacua thus sets the expected scale of baryon-violating processes, which is an observable that could potentially be probed by experiments, either at colliders \cite{Aoyama:1986ej,Ringwald:1989ee,Espinosa:1989qn,Farrar:1990vb,Ringwald:1990qz,Gibbs:1994cw} or cosmic ray and neutrino detectors \cite{Morris:1991bb,Morris:1993wg,Han:2003ru,Ahlers:2005zy,Anchordoqui:2005ey,Fodor:2003bn}. Getting accurate predictions for the rates of baryon-number-violating interactions is a difficult problem, due to a possible breakdown of
the semiclassical  expansion used to compute vacuum transitions. There have been extensive discussions in the literature (see for example \cite{McLerran:1989ab,Cornwall:1990hh,Arnold:1990va,Khlebnikov:1990ue,Porrati:1990rk,Khoze:1990bm,Khoze:1991mx,Rubakov:1992ec,Bezrukov:2003er,Ringwald:2003ns,Tye:2015tva}), which has not led to a definite consensus. Recent estimates point towards rates that  could be probed by future experiments \cite{Ringwald:2003ns,Tye:2015tva}. However, these estimates use different methods than previous calculations giving more negative results, and a detailed understanding of the reasons for the discrepancies is still lacking. For recent analyses of measurement prospects at colliders, cosmic ray and neutrino detectors, see for example \cite{Ellis:2016ast,Brooijmans:2016lfv,Ellis:2016dgb}.

Aside from determining the rate of observable baryon-violation effects, it should be noted that the energy barrier between topological vacua can also play a crucial role in potential explanations of the baryon asymmetry of the Universe. In the early Universe, finite temperature effects become important and affect the height of the barrier. At temperatures at which the electroweak symmetry is restored, the barrier effectively disappears and vacuum transitions are unsuppressed \cite{Arnold:1987mh,Khlebnikov:1988sr,Dine:1989kt}, while below the electroweak phase transition the tunneling rate becomes Boltzmann suppressed. In scenarios of 
electroweak baryogenesis \cite{Kuzmin:1985mm} (for reviews, see \cite{Trodden:1998ym,Morrissey:2012db}), the baryon asymmetry is created during the nucleation of bubbles of the broken electroweak phase in a first order transition, in such a way that unsuppressed vacuum transitions in the unbroken phase convert a chiral asymmetry into net baryon number. The latter can then survive in the broken phase only if the corresponding vacuum transitions are strongly suppressed, which enforces a bound on the relative size of the energy barrier with respect to the temperature  at the onset of bubble nucleation. On the other hand, in mechanisms of leptogenesis \cite{Fukugita:1986hr} (see \cite{Buchmuller:2004nz,Davidson:2008bu} for  reviews), out-of equilibrium decays or oscillations of heavy neutrinos generate a net lepton asymmetry, which is then partly reprocessed into baryon number by vacuum transitions. A viable mechanism then requires the lepton asymmetry to be generated while vacuum transitions are still active.

The existence of a minimum  energy barrier between vacua can be inferred from topological arguments \cite{Manton:1983nd}, and indeed one can calculate the field configurations at the top of the barrier. These are the so called sphalerons,
which correspond to saddle-points of a bosonic  energy functional. This functional depends on the spatial derivatives of the gauge and scalar fields, as well as the scalar potential. The resulting sphaleron configurations
involve a nontrivial profile for the scalar fields, which probe field values beyond the minimum of the scalar potential. Thus the resulting sphaleron energy is potentially sensitive to the details of the potential away from the Higgs vacuum. On the other hand, non-standard derivative interactions can also affect the energy functional and the sphaleron barrier.

The previous considerations motivate us to calculate the sphaleron barrier in nonstandard realizations of the Higgs vacuum, in order to look for possible deviations with respect to the SM value coming from 
a modified potential and/or derivative terms.  Sphaleron configurations have been calculated not only for the Standard Model \cite{Dashen:1974ck,Klinkhamer:1984di} (with a resulting energy barrier of the order of 9 TeV for the observed value of the Higgs mass), but also in a number of extensions of the Standard Model 
involving an elementary Higgs and other scalars \cite{Kastening:1991nw,Enqvist:1992kd,Bachas:1996ap,Moreno:1996zm,Kleihaus:1998bh,Grant:1998ci,Grant:2001at,Funakubo:2005bu,Ahriche:2007jp,Ahriche:2009yy,Ahriche:2014jna}. In many of these models, the deviations from the SM behaviour arise mainly due to the existence of additional scalars with electroweak charges, all of them acquiring nontrivial profiles in the sphaleron configuration. Still, the sphaleron barrier was never found to deviate substantially from its SM value. In this work, we restrict to models with a single electroweak scalar, and focus on possible 
large deformations of the SM case, either through sizable interactions that change the shape of the potential for an elementary Higgs, or by considering composite Higgs models, in which not only the potential is modified, but there are also new derivative interactions.

In the first case, a good example of a potential which is very different from that of the SM is one 
in which the Higgs vacuum is separated from the origin by a potential energy barrier at zero temperature. Such type of scenarios was introduced in reference \cite{Grojean:2004xa}, using higher-dimensional operators, and motivated by electroweak baryogenesis. A UV completion involving extra scalars with strong couplings to the Higgs was found in \cite{Espinosa:2007qk,Espinosa:2008kw}, and the large couplings were shown not to spoil perturbation theory in \cite{Tamarit:2014dua}. 
Hence, we will here adopt a general parametrization of the potential, capturing its features without worrying about the concrete realization in terms of additional scalars. We assume additional scalars to be stabilized at the origin, without inducing tadpoles in a given Higgs background, and thus playing no role in the calculation of sphaleron configurations.

Composite Higgs scenarios, well motivated by naturalness considerations, realise the Higgs boson as a pseudo-Goldstone boson with a potential that remains protected from large quantum corrections due to
an approximate global symmetry. We will center our attention on the minimal composite scenarios of reference \cite{Agashe:2004rs}, in which the pattern of global symmetry breaking is  SO(5)$\rightarrow$SO(4).

The organization of the paper is as follows. In section \ref{sec:csn} we summarize the link between B+L violating processes and the sphaleron barrier. The calculation of the sphaleron configuration in
the SM is reviewed in section \ref{sec:sphalsm}, while \ref{sec:sphaldef} focuses on the case of a deformed potential. Section \ref{sec:sphalcomp} focuses on the sphaleron energy in minimal composite Higgs models and in Section~\ref{sec:summary} we offer a summary.


\section{Overview of sphalerons and B+L violation}
\label{sec:csn}

In a nonabelian gauge theory, vacua are associated with pure gauge configurations: since the Hamiltonian is gauge invariant, such configurations have the same energy as the one with zero gauge fields. Fore more general field configurations, the requirement of finite action demands them to tend to such vacuum configurations at infinity. ``Infinity'' can be understood as a 3-sphere $S_3$ of infinite radius within $\mathbb{R}^4$, and thus finite action configurations are associated with mappings from $S_3$ to the gauge group. If the group is compact, such as the electroweak SU(2)$_L$, which itself 
has the topology of a sphere, the mappings are classified by an integer 
winding number or topological charge, counting the number of times that the compact group can be wrapped around $S_3$.

This topological charge $q$ can be written in terms of the nonabelian field-strength as
\begin{equation}
 q=\frac{1}{16\pi^2}\int d^4x\, {\rm tr} \,\tilde F_{\mu\nu}F^{\mu\nu},
\end{equation}
where $\tilde F^{\mu\nu}\equiv\frac{1}{2} \epsilon^{\mu\nu\rho\sigma}F_{\rho\sigma}$.
The integrand above is a total derivative, and thus only picks a contribution from the boundary at infinity, as expected from the fact that $q$ is associated with mappings of the sphere at infinity into the gauge group. One can always choose a so-called topological gauge in which $A_0=0$ and all the gauge field components go to zero at spatial infinity. Then the only nonzero contributions to $q$ at the boundary of $\mathbb{R}^4$ are localized at the two space slices at $t=\pm\infty$. It can then be seen that one may write
\begin{align}
\label{eq:qNS}
 q=N_{CS}(t=\infty)-N_{CS}(t=-\infty),
\end{align}
where $N_{CS}(t)$, known as the Chern-Simons number, is given in the topological gauge by the following integral over a spatial slice with fixed $t$:
\begin{align}
\label{eq:NCS}
 N_{CS}(t)=\frac{1}{16\pi^2}\int_{t}\!d^3x\,\epsilon_{ijk}\left(A^a_i \partial_j A^a_k+\frac{1}{3}\epsilon^{abc} A^a_i A^b_j A^c_k\right).
\end{align}
Although the topological charge $q$ is an integer, $N_{CS}$ is not necessarily so. The Chern-Simons number becomes an integer only when evaluated over pure gauge configurations. Note that, since arbitrary gauge configurations of finite action tend to a pure gauge transformation at infinity, the topological charge given by \eqref{eq:qNS} is indeed an integer.

We conclude that vacua can be characterized by integer values of the Chern-Simons number. This implies that there can be an 
energy barrier between configurations with integer $N_{CS}$. One can then consider paths in field space between vacuum configurations along which the height of the barrier is minimized. The field configurations at the top of this minimal
barriers are known as sphalerons, and the height of the barrier is the sphaleron energy. 

In order to be more precise about the aforementioned energy of the gauge field configurations, it can be defined, in analogy with a zero-dimensional quantum mechanics problem, from the contributions to the Hamiltonian that do not involve time derivatives. This gives a functional $V_{\rm bos}$ which in the topological gauge adopts the form
\begin{equation}
\label{eq:Vbos}
 V_{\rm bos}[A_\mu^a,\phi_i]\equiv\int\!d^3x\,\left\{\frac{1}{4g^2}F^a_{ij}F^{a}_{ij}+{\cal L}_{\rm kin,sp}^{\rm matter}[A_\mu^a,\phi_i]+V^{\rm matter}[\phi_i]\right\},
\end{equation}
where $\phi_i$ represents generic scalar fields, ${\cal L}_{\rm kin,sp}^{\rm matter}$ stands for the contributions of spatial derivatives to their kinetic terms, while $V^{\rm matter}$ denotes their potential energy density. Sphalerons correspond
to saddle points of $V_{\rm bos}$, as is intuitively clear from their role as configurations with maximal energy along minimal-barrier paths between vacua. Being extremal points of $V_{\rm bos}$, sphalerons are static solutions of the Euclidean equations of motion of the theory, i.e. satisfying
\begin{align}
 \partial_\nu\frac{\delta V_{\rm bos}}{\delta \partial_\nu A^a_\mu}-\frac{\delta V_{\rm bos}}{\delta  A^a_\mu}=0,\quad \partial_\nu\frac{\delta V_{\rm bos}}{\delta\partial_\nu \phi_i}-\frac{\delta V_{\rm bos}}{\delta\phi_i}=0.
\end{align}
As emphasized in Sec.~\ref{sec:intro}, because $V_{\rm bos}$ is sensitive to the potential energy density of the scalars and contributions involving their spatial derivatives, the sphaleron energy can vary if either of them is modified.

\begin{figure}[h]
\begin{center}
\includegraphics[width=0.49\textwidth]{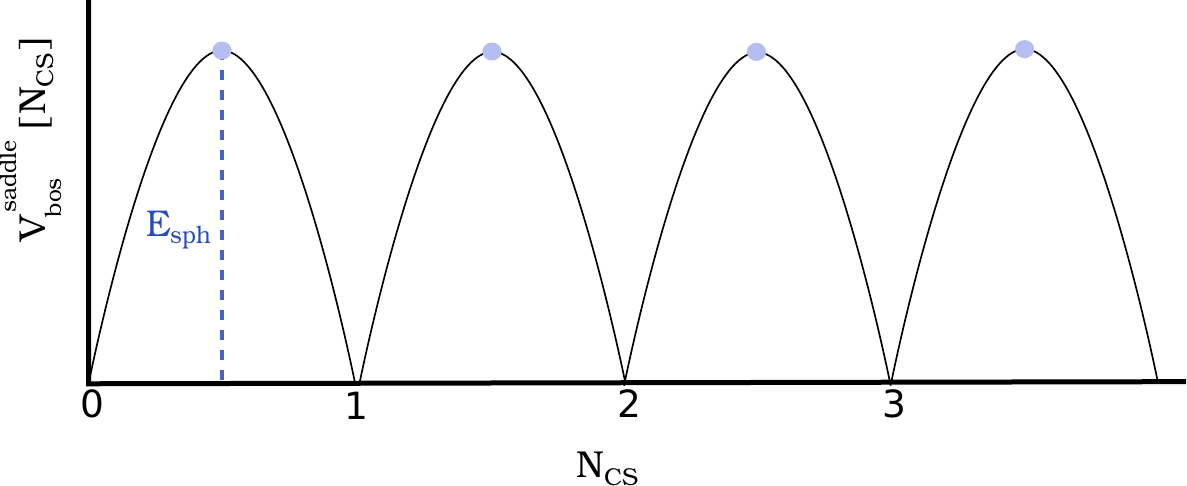}
\includegraphics[width=0.49\textwidth]{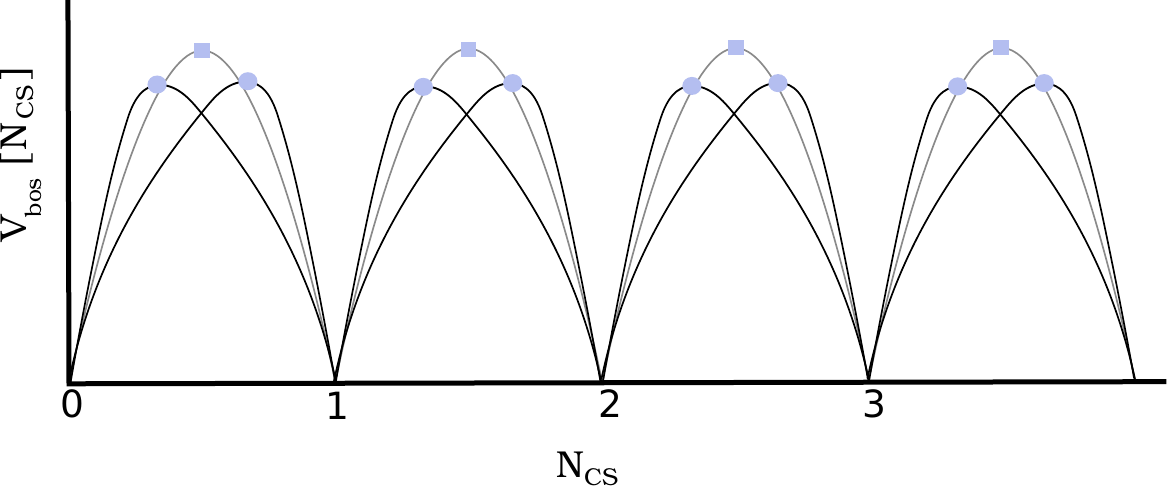}
\caption{\label{fig:VbosNCS}  Left: Schematic representation of $V^{\rm saddle}_{\rm bos}[N_{CS}]$ in the presence of a single-valued  branch of extremal solutions.  Note the translation and reflection symmetries of  the graph. Right: Illustration of   $V_{\rm bos}[N_{CS}]$ evaluated at non-extremal paths between vacua when bisphalerons are present (right).}
\end{center}
     \end{figure}

Aside from the sphaleron configurations, which are extrema of $V_{\rm bos}$, one can also define constrained extrema of $V_{\rm bos}$ by demanding a fixed value of $N_{CS}$. This gives a function $V^{\rm saddle}_{\rm bos}[N_{CS}]$. Sphalerons correspond to local maxima of $V^{\rm saddle}_{\rm bos}[N_{CS}]$. As $V_{\rm bos}$ is invariant under gauge transformations, and because gauge transformations
with nontrivial topological charge change $N_{CS}$ by integer quantities, $V^{\rm saddle}_{\rm bos}[N_{CS}]$ is a periodic function of $N_{CS}$. Further, $V_{\rm bos}$ is also invariant under parity transformations of the fields, under which $N_{CS}$ changes sign. It then follows that the graph of the function $V^{\rm saddle}_{\rm bos}[N_{CS}]$ is invariant under reflections around lines with constant half-integer and integer values of $N_{CS}$, as illustrated in Fig.~\ref{fig:VbosNCS}.  It is known that  $V^{\rm saddle}_{\rm bos}[N_{CS}]$ can be multivalued away from integer values of $N_{CS}$, implying the existence of multiple families of extrema. In the SM, for $m_h< 12~m_W$, there is a single branch, with $V^{\rm saddle}_{\rm bos}[N_{CS}]$  having negative second derivatives in between the vacua, as in the left plot of Fig.~\ref{fig:VbosNCS}. The reflection symmetry implies then that the maximum of the curve $V^{\rm saddle}_{\rm bos}[N_{CS}]$ in between integer values of $N_{CS}$ lies at half-integer values of $N_{CS}$. Or, in other words, sphaleron configurations are invariant under parity transformations for $m_h = 125$ GeV. For $m_h\geq 12~m_W$, a new branch of sphalerons appears \cite{Kunz:1988sx,Yaffe:1989ms}, which come in pairs related by parity transformations; these are known as bisphalerons. In this case the extremal path in field-space defining $V^{\rm saddle}_{\rm bos}[N_{CS}]$ becomes  multivalued, and  $V^{\rm saddle}_{\rm bos}[N_{CS}]$ develops cusps. 
Nevertheless  one can define deformed paths for which sphaleron configurations do indeed sit atop an energy barrier \cite{Kunz:1994ah}. The situation is schematically depicted in the right plot in Fig.~\ref{fig:VbosNCS}.

The former definition of bosonic potential energy, inspired by quantum mechanics, might seem ad-hoc, so that the physical meaning of $E_{\rm sph}$ needs some further clarification. In fact, it is not obvious to see how $E_{\rm sph}$ may play a role in tunneling processes between the topological vacua. The reason is that tunneling rates are computed from solutions to the full Euclidean equations of motion, known as instantons \cite{Belavin:1975fg,'tHooft:1976fv}. These differ from sphalerons 
because the latter are static solutions, while instantons depend as well on time. Despite this, the sphaleron energy can play a role when considering not just spontaneous vacuum transitions, but scattering processes at a fixed energy. The existence of multiple topological vacua affects the wave function of the true vacuum, and this effect can be incorporated in a  path integral formalism by including sums over field configurations 
around instanton backgrounds. This gives rise to new effective instanton vertices that can be incorporated in diagrammatic expansions, which encode the nontrivial effects of the vacuum transitions. In principle, these vertices are suppressed by exponential factors involving the Euclidean action of the instantons, $\exp(-S^E_{\rm inst})$, which  as said before also determine the tunneling rates. Actual calculations show that when the external
particles have energies of the order of $E_{\rm sph}$,  the exponential suppression of the instanton effects can be lifted \cite{ McLerran:1989ab,Cornwall:1990hh,Arnold:1990va,Khlebnikov:1990ue,Porrati:1990rk,Khoze:1990bm,Khoze:1991mx,Rubakov:1992ec,Bezrukov:2003er,Ringwald:2003ns,Tye:2015tva}. Thus, $E_{\rm sph}$ can indeed be interpreted as a physical energy barrier between topological vacua, because the effect of vacuum transitions becomes unsuppressed when one prepares states with $E>E_{\rm sph}$.

A more direct connection between sphalerons and energy barriers can be established at finite temperature. Thermal fluctuations allow states with energies above the barrier, which can then induce classical 
vacuum transitions. The thermal transition rate is determined from static solutions to the Euclidean equations of motion -- i.e. sphalerons -- and the rate scales as $\exp(-S^{E,3D}_{\rm sph}/T)=\exp(-E_{\rm sph}/T)$, where $S^{E,3D}_{\rm sph}$ is the thermal Euclidean action, defined as the spatial integral of the Euclidean Lagrangian evaluated on time-independent configurations. Thermal fluctuations induce
excitations with average energy of the order of $T$, and when $T\gtrsim E_{\rm sph}$ the rate becomes unsuppressed. Again, $E_{\rm sph}$ can be interpreted as an energy barrier between the topological vacua.

We can conclude this section by reviewing the link between sphalerons and  B+L violation. In the SM, B-L is conserved while B+L is an anomalous symmetry. Denoting the SU(2)$_L$ field strength as $W_{\mu\nu}$, the B+L
current satisfies the following anomalous conservation equation,
\begin{align}
\label{eq:anomeq}
 \partial_\mu J^\mu_{B+L}=\frac{3}{8\pi^2}\,{\rm tr}\, \tilde W_{\mu\nu}W^{\mu\nu}.
\end{align}
This means that a given gauge field background with topological charge $q$ induces the following change of B+L between $t=-\infty$ and $t=\infty$:
\begin{align}
\label{eq:anomaly}
 \Delta(B+L)=\int d^3x \left[J^0_{B+L}(t=\infty)-J^0_{B+L}(t=-\infty)\right]=\int d^4 x \,\partial_0 J^0_{B+L}= \frac{3}{8\pi^2}\int d^4x\,{\rm tr}\, \tilde W_{\mu\nu}W^{\mu\nu}=6 q,
\end{align}
where we used Eq.~\eqref{eq:anomeq} with the assumption that the current vanishes at spatial infinity. Tunneling between topological vacua is associated with instanton configurations 
which tend towards pure gauge configurations with different integer values of $N_{CS}$ at $t=\pm \infty$. Thus the instanton configurations have a nonzero topological charge $q=N_{CS}(\infty)-N_{CS}(-\infty)$, which implies that vacuum transitions are immediately associated with violations of B+L. In this way, the sphaleron energy sets the scale of baryon-number-violating processes. Equation  \eqref{eq:anomaly} implies that in a vacuum transition with $\Delta N_{CS}=1$, there is a change of B+L  by six units. Thus, sphaleron-related processes  involve the production of large numbers of particles. The allowed processes can be identified by using the 
effective instanton vertices mentioned earlier. For an instanton background with topological charge $q$, the vertices involve a number of fermion fields related to the number of fermionic zero modes of the background; the resulting interaction  violates B+L by $6q$ units. For example, a one-instanton vertex inducing a transition with  $\Delta N_{CS}=1$, generates an interaction with twelve fermion fields, of the form $\Pi_i (u_L d_L d_L \nu_L)_i,$ with $i=1,\dots 3$ labelling the generations \cite{Harvey:1990qw}. This can for example give rise to the creation of three baryons and three neutrinos from the vacuum, or can induce $2\rightarrow  10$ processes with quarks and leptons. As mentioned before, the production cross sections are up for debate.

In the following, we will  calculate the sphaleron energy in elementary Higgs boson scenarios with a modified potential, and in composite Higgs boson scenarios. In the first case, the modified Higgs potential can be understood
as arising from the virtual effects of heavier fields. In the second case, the sphaleron energy can be calculated in an effective
theory arising after integrating out modes of the strongly coupled sector. We have argued before that sphaleron effects become relevant at processes with energies of the order of the sphaleron energy. Then if $E_{\rm sph}$ is larger than the mass of the  heavy fields or the compositeness scale, the question might arise of whether at those energies one can still trust the original calculation of the minimum energy barrier. This is the case because the effective theory in which the heavy fields are integrated out describes the dynamics when those fields lie at their energy minima, and so minimal energy configurations of the full theory can be reliably calculated in the effective description. In the composite case, it should be noted that the Higgs, being a pseudo-Goldstone boson, is protected by the global symmetry of the composite sector. Interactions inside the latter  cannot generate contributions to the Higgs potential, which arises from interactions that break the global symmetry and are already taken into account in the effective theory. The situation is then similar to the case of an elementary Higgs with an effective  potential induced by heavy fields, and the previous conclusion applies.


\section{Sphaleron energy in the Standard Model}
\label{sec:sphalsm}

In this section we review the calculation of the SM sphaleron configuration, mostly following the treatment in \cite{Schaldach}. As we are considering the minimum barrier between vacua with different weak topological charge, we can simply restrict to field
trajectories connecting the vacua without exciting degrees of freedom that do not couple to the weak bosons -- doing otherwise would just give higher energy configurations. This allows to ignore gluons, and forces
to consider the Higgs field. As in a nonzero Higgs background the weak bosons mix with the hypercharge boson, in principle one should take it into account it as well, but because the mixing is small, the effect is subleading (less than $1\%$, \cite{Kleihaus:1991ks,Ahriche:2014jna}) and will
be ignored. Thus one has to consider the functional
\begin{equation}
\label{eq:VbosSM}
 V^{\rm SM}_{\rm bos}[A_\mu^a,H]=\int\!d^3x\,\left\{\frac{1}{4g^2}W^a_{ij}W^{a}_{ij}+D_i H^\dagger D_i H+V(H)\right\},
\end{equation}
where $D_i H=\partial_i H-i{/2}\,\sigma^a A^a_i H$, with $\sigma^a$ being the usual Pauli matrices. $V(H)$ is the Higgs potential normalized to be zero at the Higgs vacuum, so that $V^{\rm SM}_{\rm bos}$ evaluated at the sphaleron configuration can be directly interpreted as the energy barrier between topological vacua.\footnote{This is because with this choice of normalization, the bosonic energy of the vacuum configuration with zero gauge fields and the Higgs at its VEV becomes zero.} At tree level $V(H)$  is given in terms of the Higgs mass squared $m^2_h$ and the Higgs VEV $v$ by
\begin{align}
 V(H)=-\frac{m^2_h}{2v^2}\left(H^\dagger H-\frac{v^2}{2}\right)^2.
\end{align}
It is useful to work in dimensional units, and to do so we rescale the fields and coordinates in units of the $W$ mass, which in the limit of zero Weinberg angle is $m^2_W=g^2 v^2/4$:
\begin{align}
\label{eq:rescaling}
 x^\mu&\rightarrow\frac{1}{m_W} y^\mu,& A^a_\mu&\rightarrow m_W \tilde A^a_\mu, & H&\rightarrow\frac{m_W}{\sqrt{2}g} \tilde H.
\end{align}
Then one can find the sphaleron configuration by extremising the dimensionless functional
\begin{align}
\label{eq:rescaledVbosSM}
 \tilde V^{\rm SM}_{\rm bos}=\frac{1}{g^2}\int\!d^3y\,\left\{\frac{1}{4g^2}\tilde W^a_{ij}\tilde W^{a}_{ij}+\frac{1}{2}D_i \tilde H^\dagger D_i \tilde H+\tilde V(\tilde H)\right\},
\end{align}
where $ \tilde V(\tilde H)\equiv\frac{\kappa^2}{32}(\tilde H^\dagger \tilde H-4)^2$ and $\kappa^2\equiv\frac{m^2_h}{m^2_W}$. {For simplicity we don't change the notation of the covariant derivatives, which are obtained from the usual ones by simply replacing dimensionful quantities with their dimensionless counterparts.} The equations of motion of the sphaleron configuration are
\be\begin{aligned}
\label{eq:eomSM}
 &({\cal D}_j \tilde W_{ij})^a+\frac{i}{4}\left(\tilde H^\dagger \sigma^a D_i \tilde H-D_i \tilde H^\dagger \sigma^a D_i\tilde H\right)=0,\\
 &\left[D^2_i-2\frac{\partial}{\partial ( \tilde H^\dagger \tilde H)}\tilde V(\tilde H)\right]\tilde H=0,
\end{aligned}\ee
{with $({\cal D}_k \tilde W_{ij})^a=\partial_k\tilde W_{ij}^a+\epsilon^{abc}\tilde A_k^b \tilde W_{ij}^c$.}
To solve the former equations, we impose a rotationally symmetric ansatz\footnote{This ansatz is often called hedgehog solution.} \cite{Dashen:1974ck,Manton:1983nd,Klinkhamer:1984di,Akiba:1988ay}. Defining $r\equiv \sqrt{\sum y^2_i}$ and $n_i\equiv y_i/r$, the ansatz is given by
\begin{equation}
\label{eq:ansatz}
\begin{aligned}
&\tilde  A_i^a=\epsilon_{aij}n_j\frac{1-A(r)}{r}+(\delta_{ai}-n_an_i)\frac{B(r)}{r}+n_an_i\frac{C(r)}{r},\\
 &\tilde H(r)=2\left(F(r)\,\mathbb{I}+i G(r)\,\vec{n}\cdot\vec{\sigma}\right)\left[
 \begin{array}{c}
  0\\
  1
 \end{array}\right].
 \end{aligned}
\end{equation}
One can consider SU(2)$_L$ transformations preserving the $A_0=0$ gauge condition. Taking a group element of the form
\begin{equation}
 U(r)=\exp[{\vec{n}\cdot\vec{\sigma}} P(r)]=\cos P(r)+i{\vec{n}\cdot\vec{\sigma}}\sin P(r),
\end{equation}
the functions in the ansatz of \eqref{eq:ansatz} transform as
\begin{equation}
\label{eq:gaugetr}
\begin{aligned}
 A\rightarrow &~ A\cos 2P-B \sin 2P,& B\rightarrow&~ B\cos 2 P+ A\sin 2P, &C\rightarrow&~ C+ 2r P',\\
 F\rightarrow &~ F \cos P - G\sin P, & G\rightarrow&~ G\cos P+F \sin P.
 \end{aligned}
\end{equation}
We can use this freedom to set $C(r)=0$, although the price to pay is that one will lose the topological gauge condition $A_i\rightarrow0$ for $r\rightarrow\infty$.\footnote{In the topological gauge, given the ansatz \eqref{eq:ansatz}, $A_i\rightarrow0$  implies for example $A(r)\rightarrow1$, which is not respected by the gauges transformations of equation \eqref{eq:gaugetr}.} Inserting the ansatz \eqref{eq:ansatz} with $C(r)=0$ into the first of the equations in \eqref{eq:eomSM}, one gets an equation of the form
\begin{equation}
\label{eq:eqgauge}
 \begin{aligned}
  E_1 \frac{2n_an_i}{r^2}+E_2\frac{n_an_i-\delta_{ai}}{r}+E_3\,\epsilon_{aij}\frac{n_j}{r}=0.
 \end{aligned}
\end{equation}
The orthogonality of the 2-index objects with indices $a,i$ of Eq.~(\ref{eq:eqgauge}) means that its
solutions must satisfy $E_1=E_2=E_3=0$, which yields
\begin{equation}
\label{eq:Es}
 \begin{aligned}
 &\,B A'-AB'+r^2 (GF'-FG')=0,\\
 &\,B''-\frac{B}{r^2}\left(A^2+B^2-1\right)+2GF-B(G^2+F^2)=0,\\
 &\,A''-\frac{A}{r^2}\left(A^2+B^2-1\right)-A(G^2+F^2)-G^2+F^2=0.
 \end{aligned}
\end{equation}
The second equation in \eqref{eq:eomSM}, after substitution of the ansatz, adopts the form 
\begin{align}
 E_4\,\mathbb{I}+E_5\,{\vec{n}\cdot\vec{\sigma}}=0.
\end{align}
This implies $E_4=E_5=0$, which gives
\be\label{eq:E4E5SM}\begin{aligned}
 &\frac{2}{r^2}(r^2G')'-\frac{G}{r^2}\left((A+1)^2+B^2\right)+\frac{2BF}{r^2}-\kappa^2G(F^2+G^2-1)=0,\\
 &\frac{2}{r^2}(r^2F')'-\frac{F}{r^2}\left((A-1)^2+B^2\right)+\frac{2BG}{r^2}-\kappa^2F(F^2+G^2-1)=0.
\end{aligned}
\ee
By calculating the derivative with respect to $r$, one can show that the first equation of Eqs.~\eqref{eq:Es} is not independent of the others, leaving four equations with four unknown functions.
 Solving them requires to impose boundary conditions for the unknown functions and their derivatives. 
 At large $r$, finiteness of $V_{\rm bos}$ evaluated with the sphaleron solution implies that gauge fields must approach a pure gauge configuration, while the scalar fields must tend to a minimum of their potential. The choice of boundary conditions can be simplified by obtaining asymptotic solutions with the desired properties, 
 which will depend on fewer parameters. For the SM, the asymptotic solutions for large and small $r$ at the chosen accuracy level depend each on 3 parameters, and are given in appendix \ref{app:asympt}. 
 
  A regular sphaleron solution can be found by applying an iterative numerical procedure such that, at each step, one obtains two solutions to the sphaleron equations by imposing boundary conditions at large and small r, respectively, while the steps are repeated with varying boundary conditions until the two solutions match smoothly at an intermediate value of $r$.

Before illustrating the solutions, it should be noted that one can reduce the equations further by redefining the unknown functions. Given
the gauge transformation properties \eqref{eq:gaugetr}, one may define gauge-invariant quantities $R^2\equiv A^2+B^2$, $S^2\equiv {F}^2+G^2$. Then one has
\begin{equation}
\label{eq:RS}
 \begin{aligned}
  A=&R \cos\theta, & B=& R \sin\theta,\\
  F= &\,S \cos \phi, & G=& S\sin\phi.
 \end{aligned}
\end{equation}
The above mapping does not uniquely define the variables $R,S,\theta,\phi$, since  $A,B,F,G$ are invariant under two discrete transformations, i.e. 
\be\begin{aligned}
\label{eq:jumps1}
 R\rightarrow R,\quad \theta=\theta+2m\pi, \\
 S\rightarrow S,\quad \phi\rightarrow\phi+2n\pi,
\end{aligned}\ee
and
\be\begin{aligned}
\label{eq:jumps2}
 R\rightarrow-R,\quad \theta=\theta+(2m+1)\pi, \\
 S\rightarrow-S,\quad \phi\rightarrow\phi+(2n+1)\pi,
\end{aligned}\ee
with $m,n,\in\mathbb{Z}$.
 When looking for smooth sphaleron profiles, it should be noted that the former discrete changes in $R,S,\theta,\phi$ can still be admitted, since they don't affect
 the functions $A,B,F,G$. In terms of the new variables the four independent equations become
\be
\label{eq:RSSM}
\begin{aligned}
&r^2R''+r^2S^2 \cos [2 \phi -\theta ]+R-R \left(R^2+r^2(\theta '^2+S^2\right))=0,\\
&2 r^2 S''-2r^2 S \phi'^2+4r S'-S \left(\kappa^2 r^2 \left(S^2-1\right)-2 R \cos [2 \phi -\theta ]+R^2+1\right)=0,\\
&R \theta ''+2 \theta ' R'+S^2 \sin [2 \phi -\theta ]=0,\\
&r^2 S \phi''+2 r\phi' \left(r S'+S\right)-R S \sin [2 \phi -\theta ]=0.
\end{aligned}\ee
The last two equations can be solved by 
\begin{align}
\label{eq:thetas}
 \theta'=\phi'=0~~\mathrm{and}~~\phi=\frac{\theta}{2}+\omega\frac{\pi}{2},\,~~\mathrm{with}~~\omega\in \mathbb{Z},
\end{align}
which finally yields
\begin{equation}
\label{eq:eqsRsimpleSM}
 \begin{aligned}
 &r^2 R''-{R^3}+ R \left(1-r^2 S^2\right)\pm r^2 S^2=0,\\
 &2r^2 S''+4 r S'-S \left((R\mp1)^2+\kappa^2 r^2(S^2-1)\right)=0.
 \end{aligned}
\end{equation}
The upper and lower signs are associated with even and odd $\omega$ in \eqref{eq:thetas}, and the corresponding equations can be related by the transformation $R\rightarrow-R$. However, if the sign of $R$ is fixed at large values of $r$ with a suitable boundary condition, both types of equations could give rise to different branches of sphalerons. For $m_h=125$ GeV and $R>0$ at large values of $r$, only the upper-sign branch has solutions. Equations \eqref{eq:thetas}, \eqref{eq:eqsRsimpleSM} do not allow to fix the constant values of $\theta,\phi$, which, given the identities in Eq.~\eqref{eq:RS}, prevents to reconstruct the values of the four unknown functions $A,B,{F},G$ in the ansatz \eqref{eq:ansatz} in the gauge $C=0$. Nevertheless, $\theta$ can be determined from the generic properties of the functional $V^{\rm saddle}_{\rm bos}[N_{CS}]$ introduced in section \ref{sec:csn}, up to the ambiguity of Eqs.~\eqref{eq:jumps1} and \eqref{eq:jumps2}. As mentioned before, for the observed value of the Higgs mass there is a single branch of parity-invariant sphaleron solutions, and the symmetries 
of $V^{\rm saddle}_{\rm bos}[N_{CS}]$  then imply that sphalerons have $N_{CS}=1/2+n,\,\, n\in \mathbb{Z}$. In order to get the expression of $N_{CS}$ in the $R,S,\theta,\phi$ 
field coordinates, one has to be careful because the relation of Eq.~\eqref{eq:NCS} for $N_{CS}$ is only valid in a topological gauge with $A_i\rightarrow0$ for $r\rightarrow\infty$. However, in order to eliminate
the function $C(r)$ from the ansatz \eqref{eq:ansatz} we performed a further gauge transformation which can violate the previous gauge condition. Nevertheless, one can use the properties of gauge transformations in Eq.~\eqref{eq:gaugetr} to map the fields in the $C=0$ gauge into fields in the topological gauge, where Eq.~\eqref{eq:NCS} holds. Expressing the result in terms of functions in the $C=0$ gauge one finally obtains:
\begin{equation}
\label{eq:NCSC0}
 N_{CS}=\frac{1}{2\pi}\int dr (A'B-B'A)+\frac{1}{2\pi}\arctan\frac{B_\infty}{A_\infty}=\frac{\theta_\infty+ n\pi}{2\pi}-\frac{1}{2\pi}\int dr R^2\theta', \,\,n\in\mathbb{Z}.
\end{equation}
$B_\infty, A_\infty,\theta_\infty$ denote  the values of the corresponding functions at infinity. In the $C=0$ gauge it no longer holds that the gauge fields vanish at $r\rightarrow\infty$. The ambiguity in $\theta_\infty$ up to multiples of $\pi$ is due to the discrete redundancy of Eqs.~\eqref{eq:jumps1} and \eqref{eq:jumps2}. From Eq.~\eqref{eq:NCSC0}, when imposing Eq.~\eqref{eq:thetas} one can see that the sphaleron solutions with $N_{CS}=1/2$ have constant $\arctan B_\infty/A_\infty=\theta_\infty+ n\pi=\pi$. This, together with Eq.~\eqref{eq:thetas}, allows to fix the ansatz  \eqref{eq:ansatz} in the $C=0$ gauge by simply solving the two differential equations in \eqref{eq:eqsRsimpleSM}.
The boundary conditions for the two functions $R$ and $S$ can be obtained from the asymptotic solutions for the functions $A,B,G,F$ in appendix \ref{app:asympt}, imposing Eq.~$\eqref{eq:thetas}$ and $\theta=\pi+n\pi$.
At the chosen level of accuracy, this reduces the free parameters of the asymptotic solutions from six to four.

For $N_{CS}=1/2$ and $R>0$ at large values of $r$,  only the upper sign choice in Eq.~\eqref{eq:eqsRsimpleSM} gives a solution, and one can choose  $\theta=\pi$. The upper sign choice  corresponds to even $\omega$ in Eq.~\eqref{eq:thetas}, i.e.  $\phi=\theta/2+n \pi,$ with $n\in\mathbb{Z}$. As is clear from Eq.~\eqref{eq:RS}, this implies that the sphaleron has $F=B=0$ for all $r$.  As mentioned earlier,  for $r\rightarrow\infty$ the scalar field must lie in a minimum of its potential energy  in order for the sphaleron to have finite energy. This is satisfied for $F^2+G^2=S^2=1$, as can be seen from the ansatz \eqref{eq:ansatz} and the rescaled potential term in $\tilde V_{\rm bos}^{\rm SM}$ in Eq.~\eqref{eq:rescaledVbosSM}.  On the other hand, regularity at $r=0$ forces $G(0) = 0$, which, together with the condition
$F(r)=0\,\forall\, r$, means that the scalar field must be zero at $r=0$. Thus, the sphaleron probes the Higgs potential between the origin ($F=G=0$) and the vacuum configuration ($F^2+G^2=1$).

The sphaleron energy can be obtained from $m_W \tilde V_{\rm bos}$ evaluated in the sphaleron configuration; in terms of the $R,S,\theta,\phi$ variables, $\tilde V_{\rm bos}$  is equal to
\be\begin{aligned}
\label{eq:tildeVbosSM}
\tilde V_{\rm bos}=\frac{2\pi }{g^2}\int \frac{dr}{r^2}\,&\left\{2 r^2 \left[{R'}^2+R^2 {\theta '}^2+2 r^2 \left(S^2 {\phi '}^2+{S'}^2\right)\right]+\kappa ^2 r^4 \left(S^2-1\right)^2+2 R^2 \left(r^2 S^2-1\right)\right.\\
 &\left.-4 r^2 R S^2 \cos [\theta -2 \phi ]+2 r^2 S^2+R^4+1\right\}.
\end{aligned}\ee

\begin{figure}[h!]
\begin{center}
\begin{minipage}{0.5\textwidth}
 \includegraphics[width=.95\textwidth]{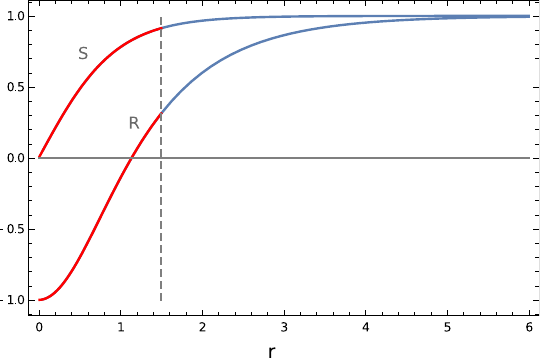}
\end{minipage}%
\begin{minipage}{0.5\textwidth}
\hfil\hskip0.8cm\includegraphics[width=.95\textwidth]{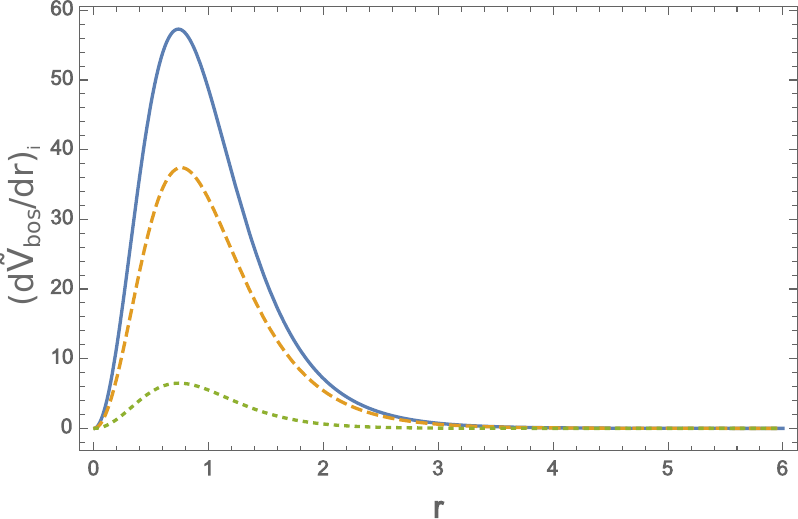}
\end{minipage}
\caption{Left: Profiles for $R,S$, in units of $m_W$, in the SM sphaleron configuration obtained by solving the reduced system of 2 differential equations. The vertical line marks the scale at which the low $r$ solution (red) was matched with the high $r$ solution (blue). Right: Contributions to the dimensionless integrand in $\tilde V_{\rm bos}^{SM}$, evaluated on the sphaleron solution, due to the gauge fields (solid blue), derivatives of the scalar field (dashed orange) and the potential
energy density of the Higgs (dotted green).}
\label{fig:profiles_SM}       
\end{center}
\end{figure}

Solving the different  systems of equations -- either \eqref{eq:Es} and \eqref{eq:E4E5SM}, or the system \eqref{eq:RSSM}, or the reduced system \eqref{eq:eqsRsimpleSM} -- with the iterative procedure described above, fixing $m_h=125.09$ GeV and $m_W=80.398$ GeV \cite{Olive:2016xmw} we recover in all cases the known value of the SM sphaleron barrier,
\begin{align}
 E^{SM}_{\rm sph}=9.11\, {\rm TeV}.
\end{align}
Fig. \ref{fig:profiles_SM} illustrates the profiles for $R$ and $S$ in the sphaleron solution, as well as the contributions to the dimensionless bosonic energy density -- defined as the integrand in equation \eqref{eq:tildeVbosSM} -- from the derivatives of the gauge fields, those of the scalars, and the scalar potential. The contribution from the potential is substantially lower than that of the derivatives. This hints towards a limited sensitivity of $E_{\rm sphal}$ to the details of the scalar potential, and greater sensitivity to modified derivative interactions.  This will be confirmed in the following section dedicated to nonstandard Higgs scenarios.


\section{Sphaleron energy for an elementary Higgs in a deformed potential}
\label{sec:sphaldef}
As an illustration of the effect of a modified potential away from the Higgs vacuum, in this section we consider a theory with an elementary Higgs, yet with a nonstandard potential. The experience with the SM shows that the sphaleron configuration for an elementary Higgs is sensitive to field values between the origin and the vacuum configuration, as follows from the 
boundary conditions at $r\rightarrow\infty$ and $r\rightarrow0$. Thus we may consider potentials which deviate from the SM in this
region, while having a minimum whose VEV and curvature reproduce the correct Higgs and gauge boson masses. A potential which is very different from the SM can be achieved for example if the Higgs vacuum at zero
temperature is 
separated from the origin of field space by a potential energy barrier. Such type of scenarios was introduced in reference \cite{Grojean:2004xa}, using higher-dimensional operators. A UV completion involving
extra scalars with strong couplings to the Higgs boson was found in \cite{Espinosa:2007qk,Espinosa:2008kw}, and the large couplings were shown not to spoil perturbation theory in \cite{Tamarit:2014dua}. Here we will adopt a practical
approach and simply model the Higgs potential with strong logarithmic corrections, i.e.
\begin{align}
\label{eq:VHdef}
 V(H)=V_0+m_H^2H^\dagger H+(H^\dagger H)^2\left(-\lambda+\beta\log\left[\gamma+\frac{2H^\dagger H}{\phi^2_0}\right]\right).
\end{align}
In the equation above, $\beta$ represents an effective quartic coupling arising from loop corrections, and $\gamma$ -- which would be associated with the field-independent contributions to the masses of the
particles running in the loop corrections -- guarantees that the potential is analytic at $H=0$. $\phi_0$ can be chosen at will to be the Higgs VEV $v$ (the difference can be compensated by a redefinition
of the other couplings), and $V_0$ is fixed by requiring as before that the potential is zero at the minimum. Imposing that the correct Higgs and $W$ masses are generated at tree-level, one can eliminate the couplings $m^2_H$ and $\lambda$, and end up with a potential
\begin{align}
\label{eq:VHdef2}
 V^{\rm log}(H)=H^\dagger H\left(-\frac{m^2_h}{2}+\frac{(2+3\gamma)\beta v^2}{2(1+\gamma)^2}\right)+(H^\dagger H)^2\left(\frac{m^2_h}{2v^2}-\frac{\beta(3+4\gamma)}{2(1+\gamma)^2}+
 \beta\log\left[\frac{\gamma v^2+2H^\dagger H}{v^2(1+\gamma)}\right]\right),
\end{align}
 with only $\beta$, $\gamma$ as free parameters.
 
\begin{figure}[h!]
\begin{center}
\begin{minipage}{0.5\textwidth}
 \includegraphics[width=0.95\textwidth]{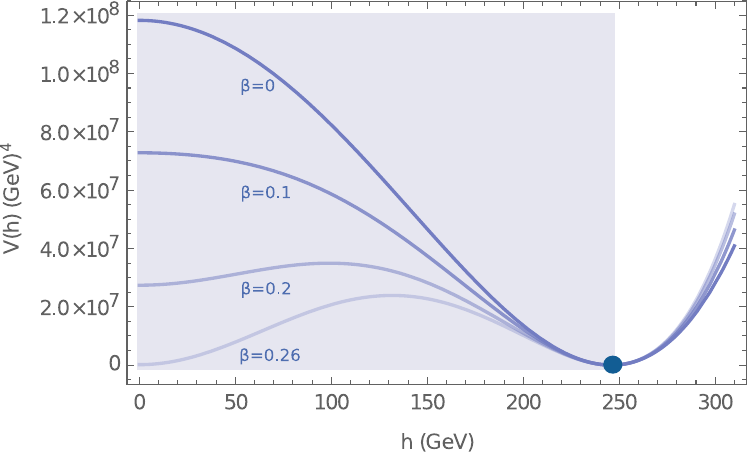}
\end{minipage}%
\begin{minipage}{0.5\textwidth}
\includegraphics[width=0.95\textwidth]{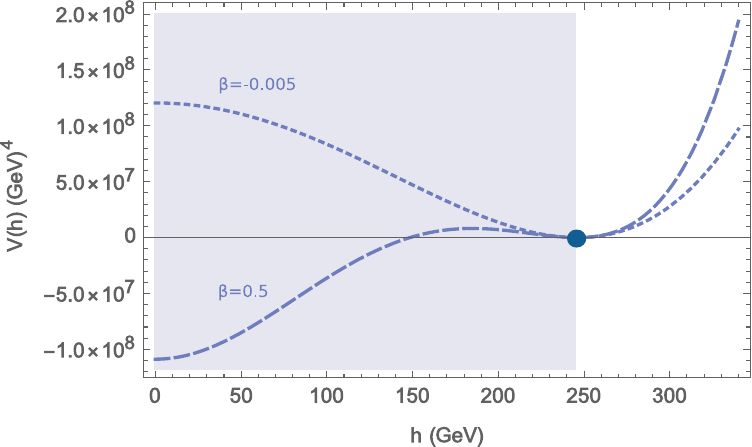}
\end{minipage}
\caption{Deformed Higgs potential for $\gamma=0.1$ and varying values of $\beta$. The dots represent the Higgs minimum, with VEV and curvature fixed by the $W$ and Higgs masses. On the left hand, the values of $\beta$ ensure absolute stability of the Higgs vacuum. The right
hand plot shows the extremal values of $\beta$ for which the metastable Higgs vacuum is still sufficiently long lived at zero temperature. The shaded areas reflect the region of the potential probed by sphaleron configurations.}
\label{fig:potentialdef}       
\end{center}
\end{figure}

\begin{figure}[h!]
\begin{center}
 \includegraphics[width=.5\textwidth]{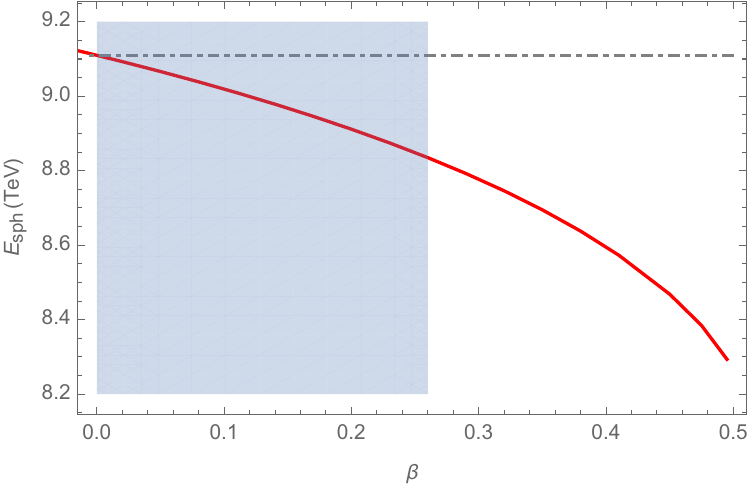}
\caption{In red, sphaleron energy as a function of $\beta$ in models with a deformed Higgs potential, for $\beta$ in the allowed stability window. The dash-dotted gray line represents the SM result. $\gamma$ was fixed to $0.1$, and hardly influences the results. The shaded band corresponds to absolutely stable Higgs vacua, as in the left plot in figure \ref{fig:potentialdef}.}
\label{fig:Esph_def}       
\end{center}
\end{figure}

\begin{figure}[h!]
\begin{center}
\begin{minipage}{0.5\textwidth}
 \includegraphics[width=.95\textwidth]{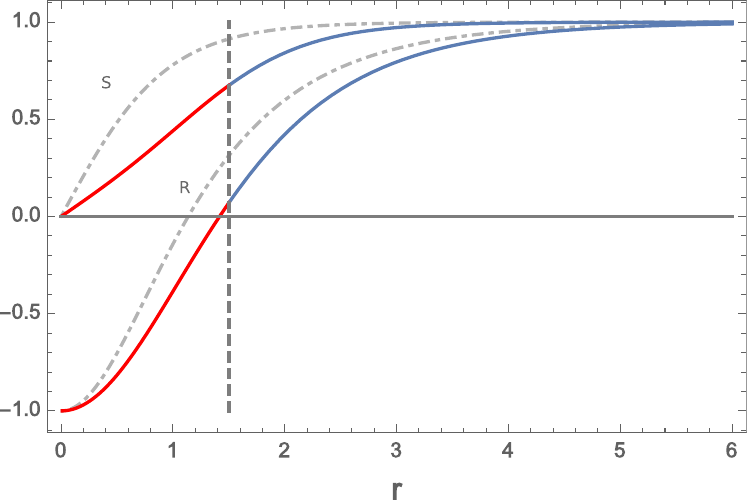}
\end{minipage}%
\begin{minipage}{0.5\textwidth}
\hfil\hskip0.8cm\includegraphics[width=.95\textwidth]{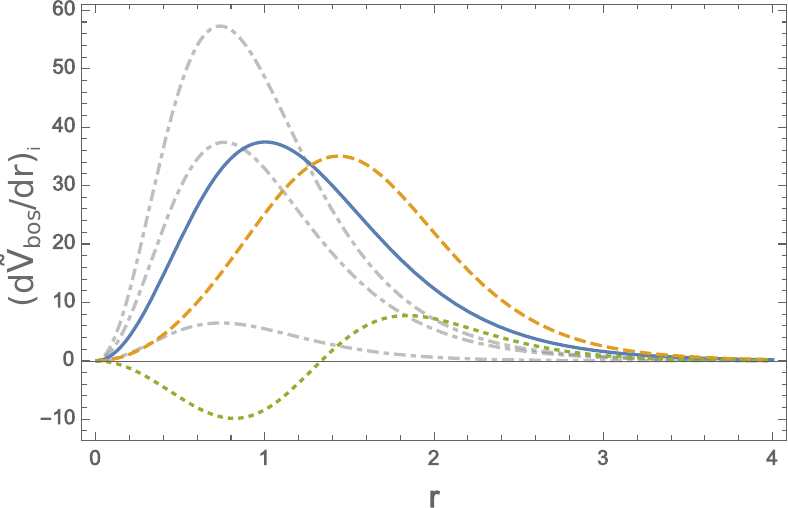}
\end{minipage}
\caption{Properties of sphaleron configurations with  $\gamma=0.1$ and $\beta=0.495$. The left plot shows the profiles for $R,S$, in units of $m_W$, with the vertical lines marking the scale at which the low $r$ solution (red) was matched with the high $r$ solution (blue). The plot on the right shows the contributions to the dimensionless integrand in $\tilde V_{\rm bos}$, evaluated on the sphaleron solution, due to the gauge fields (solid blue), derivatives of the scalar field (dashed orange) and the potential
energy density of the Higgs (dotted green). In all plots, the gray dash-dotted lines correspond to the SM results.}
\label{fig:profiles_def}       
\end{center}
\end{figure}

The parameter $\beta$ controls the size of the barrier with respect to the origin and the energy of the Higgs vacuum. A barrier appears for $\beta>0$, yet increasing $\beta$ too much ($\gtrsim 0.26$ for the measured values of $m_h$ and $m_W$) raises the Higgs vacuum above the origin, so that the symmetric phase becomes preferred. A negative value of $\beta$ causes an instability at values of the field beyond the Higgs vacuum, 
which captures the situation in the SM for the measured value of the Higgs and top masses. The allowed window of values of $\beta$ can be obtained by requiring that the electroweak vacuum is sufficiently long-lived with 
respect to tunneling towards large values of the fields (for $\beta<0$) or towards the origin ($\beta>0$). The tunneling rate can be calculated from the exponential of the Euclidean action of the scalar field evaluated at a bounce solution \cite{Coleman:1977py}. We have computed the latter  numerically for $\beta>0$, while for $\beta<0$, in the presence of a runaway as in the SM, we used the analytic approximation of \cite{Isidori:2001bm}. Doing so  we obtain the following window of allowed parameters:
\begin{align}
\label{eq:stabwind}
 -0.005\lesssim\beta\lesssim0.5.
\end{align}
The shape
of the potential is illustrated in figure \ref{fig:potentialdef} for different values of $\beta$, including the extrema of the above interval.
If the extra scalar fields that originate the logarithmic corrections are stabilized at the origin and do not receive induced tadpoles in a given Higgs background (as can be ensured
with appropriate discrete or global symmetries), they will not play a role in the calculation 
of the sphaleron barrier and can be set to zero. Thus the sphaleron configuration can be obtained, as in the SM case, by extremising  the  bosonic energy functional involving the spatial 
derivatives of the gauge fields and the Higgs, and the modified Higgs potential of equation \eqref{eq:VHdef2}. Performing the same  rescalings as in Eq.~\eqref{eq:rescaling}, the rescaled bosonic energy looks as Eq.~\eqref{eq:rescaledVbosSM}, but with $\tilde V(\tilde H)$ substituted by
\begin{align}
 \tilde V(\tilde H)=\frac
 {\kappa^2}{32}(\tilde H^\dagger \tilde H-4)^2-\frac{2\hat\beta(1+2\gamma)}{(1+\gamma)^2}+\frac{\hat\beta(2+3\gamma)}{(1+\gamma)^2}\tilde H^\dagger \tilde H+\frac{\hat\beta(\tilde H^\dagger\tilde  H)^2}{4}\left\{\log\left[\frac{4\gamma+\tilde H^\dagger \tilde H}{4(1+\gamma)}\right]-\frac{(3+4\gamma)}{2(1+\gamma)^2}\right\},
\end{align}
where we defined $\hat\beta\equiv \beta/g^2$. The equations of the sphaleron are formally the same as in Eq.~\eqref{eq:eomSM}, but with the above potential. Introducing again the ansatz 
\eqref{eq:ansatz} and choosing the gauge $C(r)=0$, one gets identical results as before for the first family of equations in \eqref{eq:eomSM}, i.e. Eqs \eqref{eq:Es}, as they are not sensitive to the potential, while the second family of equations is modified. Once more, there are only  four independent equations, which are:
\begin{equation}
\label{eq:Esdef}
 \begin{aligned}
 &\,B''-\frac{B}{r^2}\left(A^2+B^2-1\right)+2GF-B(G^2+F^2)=0,\\
 &\,A''-\frac{A}{r^2}\left(A^2+B^2-1\right)-A(G^2+F^2)-G^2+F^2=0,\\
  &\frac{2}{r^2}(r^2G')'-\frac{G}{r^2}\left((A+1)^2+B^2\right)+\frac{2BF}{r^2}-\kappa^2G(F^2+G^2-1)+G\Delta=0,\\
 &\frac{2}{r^2}(r^2F')'-\frac{F}{r^2}\left((A-1)^2+B^2\right)+\frac{2BG}{r^2}-\kappa^2F(F^2+G^2-1)+F\Delta=0,
 \end{aligned}
\end{equation}
where we defined
\be
\label{eq:gammadeltasdef}
\begin{aligned}
 \Delta\equiv&\,\frac{4 \beta }{\gamma _0^2 \gamma _1}\Big\{2 \gamma _0^2 \gamma _1 \left(F^2+G^2\right)\log \frac{\gamma _0}{\gamma _1} -\left(F^2+G^2-1\right) \left(\gamma ^2 \left(F^2+G^2
 -3\right)-2 \gamma  \left(F^2+G^2+1\right)-2 \left(F^2+G^2\right)\right)\Big\},\\
 \gamma_0\equiv&\,(1+\gamma),\\
 \gamma_1\equiv&\,(F^2+G^2+\gamma).
\end{aligned}
\ee
The presence of $F, G$ in the gauge-invariant combination suggests that the equations will be simpler using the variables $R,S,\theta,\phi$ as in Eq.~\eqref{eq:RS}. We find
\be
\label{eq:eqsRdef}
\begin{aligned}
&r^2R''+r^2S^2 \cos [2 \phi -\theta ]+R-R \left(R^2+r^2(\theta '^2+S^2\right))=0,\\
&2 r^2 S''-2r^2 S \phi'^2+4r S'-S \left(\kappa^2 r^2 \left(S^2-1\right)-2 R \cos [2 \phi -\theta ]+R^2+1\right)+r^2 S\Delta_S=0,\\
&R \theta ''+2 \theta ' R'+S^2 \sin [2 \phi -\theta ]=0,\\
&r^2 S \phi''+2 r\phi' \left(r S'+S\right)-R S \sin [2 \phi -\theta ]=0,
\end{aligned}
\ee
where now, using $\gamma_0=1+\gamma$ and $\gamma_1=S^2+\gamma$ as in Eqs.~\eqref{eq:gammadeltasdef},
\be
\begin{aligned}
 \Delta_S\equiv-\frac{4\hat\beta }{\gamma_0^2\gamma_1}\left\{\gamma _1 \left(3 \gamma -(4 \gamma +3) S^2+2\right)+\gamma _0^2 S^2 \left(2 \gamma _1 \log\frac{\gamma _1}{\gamma_0}+S^2\right)\right\}.
\end{aligned}
\ee
As in the SM case, the last two equations are solved by Eq.~\eqref{eq:thetas}, and one gets a simplified set of only two differential equations:
\begin{equation}
\begin{aligned}
\label{eq:eqsRsimpledef}
  &r^2 R''-{R^3}+ R \left(1-r^2 S^2\right)\pm r^2 S^2=0,\\
 &2r^2 S''+4 r S'-S \left((R\mp1)^2+\kappa^2 r^2(S^2-1)\right)+r^2 S\Delta_S=0.
 \end{aligned}
\end{equation}
Once more, the upper and lower sign correspond to branches of solutions with either even or odd $\omega$ in Eq.~\eqref{eq:thetas}.
The asymptotic solutions and the boundary conditions for the equations are similar to those in the SM case, and discussed in appendix \ref{app:asympt}. 
Using the same iterative method, we have solved the systems
\eqref{eq:Esdef}, \eqref{eq:eqsRdef} and \eqref{eq:eqsRsimpledef}, obtaining compatible results in all cases. As in the SM, for the measured Higgs and $W$ boson masses we only found a single branch
of sphaleron solutions, which for $R>0$ at large $r$ corresponds to the upper sign choice in \eqref{eq:eqsRsimpledef}, with $N_{CS}=1/2$, and with $\theta=\pi,\phi=\pi/2$, i.e. $F=B=0$. 
The resulting energy barrier differs at the level of $\lesssim9\%$ from the SM one, even for the limiting cases in the stability window of equation \eqref{eq:stabwind}. For absolutely stable Higgs vacua, the deviations are below 3\%; these results are illustrated in Fig.~\ref{fig:Esph_def}. Figure \ref{fig:profiles_def} shows the profiles and the contributions to the bosonic energy density coming from 
a sphaleron configuration with   $\beta$ near the upper stability limit of Eq.~\eqref{eq:stabwind}. This gives the largest deviation from the SM, with the Higgs minimum above the origin (see Fig.~\ref{fig:potentialdef}). Note that with the potential normalized to zero at the former minimum, the energy density at the origin becomes negative, and the sphaleron configuration probes negative energies, as shown on the right plot in Fig.~\ref{fig:profiles_def}. This  plays a role in lowering the sphaleron energy barrier, which becomes $E_{\rm sph}[\beta=0.495]=8.29$ TeV.

\section{Sphaleron energy in composite Higgs scenarios}
\label{sec:sphalcomp}

In this section we will study sphaleron configurations in minimal composite Higgs scenarios, in which the Higgs arises as a pseudo-Goldstone boson of a global SO(5) symmetry broken down to SO(4) \cite{Agashe:2004rs}.
The pattern of symmetry breaking enforces non-standard derivative interactions for the Higgs, as well as a modified relation between the Higgs VEV and the weak boson masses. Interactions that break the global symmetry generate a nonstandard Higgs potential, which still exhibits a discrete translational symmetry. Both the modified derivative interactions and potential can affect the sphaleron energy, and we expect larger deviations
than before. This is not only due to the modified derivative interactions, but also to the modified relation between the Higgs VEV and the $W$ boson mass. In models with elementary Higgses, we saw that the sphaleron probes
the potential between the origin and the minimum. With the location and curvature of the minimum fixed by the Higgs and gauge boson masses, the potential of an elementary Higgs can only be modified by changing the depth of 
the minimum, or the shape of the potential in between the latter and the origin. In composite Higgs models there is in principle a further degree of freedom associated with the location of the minimum, as a result of the modified relation between the $W$ mass and the Higgs VEV.

\begin{figure}[h!]
\begin{center}
\begin{minipage}{0.5\textwidth}
 \includegraphics[width=0.95\textwidth]{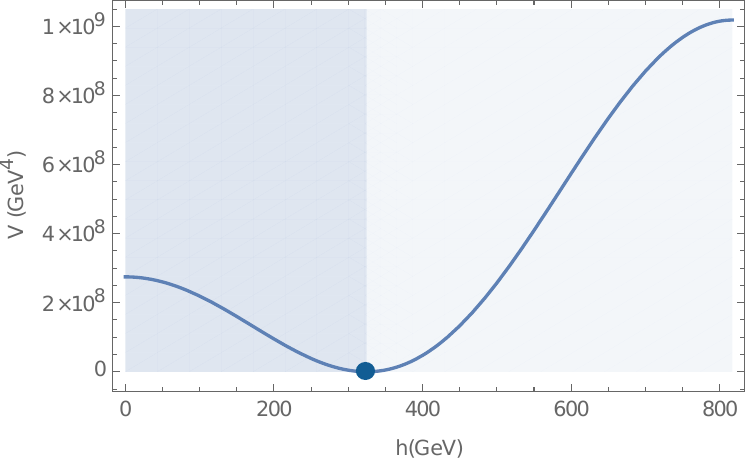}
\end{minipage}%
\begin{minipage}{0.5\textwidth}
\includegraphics[width=0.95\textwidth]{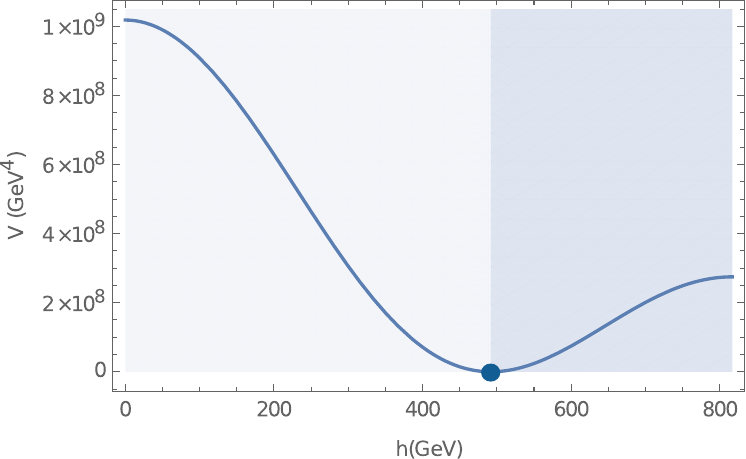}
\end{minipage}
\caption{Composite Higgs potential  for $f_\pi=260$ GeV, for the physically equivalent realizations with $\alpha<0$ (left) and $\alpha>0$ (right). Both cases yield the correct Higgs and $W$ masses at tree-level. The darker shade shows the region of the potential probed by the
sphaleron branch in common with the SM, while the lighter shade corresponds to the region probed by the new, higher-energy sphaleron branch present in composite models.}
\label{fig:Vcomp}       
\end{center}
\end{figure}

Following \cite{Agashe:2004rs}, one can work in an effective theory involving the gauge and pseudo-Goldstone fields. The breaking of SO(5) into SO(4) leaves four Goldstones $h^m,m=1,\dots,4$, which can be included
in a multiplet $\Sigma$ carrying a nonlinear representation of the broken SO(5). The breaking is assumed to originate from a field $\Sigma_0$ 
in the fundamental of SO(5), which acquires a VEV involving a scale $f_\pi$: $\langle\Sigma_0\rangle^\top=[0,0,0,0,f_\pi]$. Then  the field multiplet $\Sigma$ is given by
\begin{equation}
 \Sigma=\exp\frac{\Pi}{f_\pi}\times\Sigma_0,
\end{equation}
with $\Pi$ given by a sum over broken global SO(5) generators $\tilde G^a$ multiplied by its corresponding Goldstone fields $\Pi=i\sqrt{2} \tilde G^a h^a$. The result is
\begin{equation}
\label{eq:Sigma}
 \Sigma^\top=\frac{\sin\frac{h}{f_\pi}}{h}[h^1,h^2,h^3,h^4,h\cot\frac{h}{f_\pi}],
\end{equation}
where we defined $h\equiv(\sum_m(h^m)^2)^{1/2}$. With these conventions, and gauging an $SU(2)$ subgroup of the surviving SO(4) symmetry of the composite sector, 
the effective Lagrangian of the gauge and pseudo-Goldstone fields becomes \cite{Agashe:2004rs}
\begin{equation}
\label{eq:L}
  {\cal L}= \,\frac{f_\pi^2}{2}(D_\mu\Sigma)^\top D^\mu\Sigma-\frac{1}{4g^2}W^a_{\mu\nu}W^{a,\mu\nu}-\alpha\cos\frac{h}{f_\pi}+\beta\sin^2\frac{h}{f_\pi}.
\end{equation}
The last two terms represent the scalar potential for the pseudo-Goldstones, arising from explicit sources of SO(5) breaking, such as Yukawas and the gauging of $\mathrm{SU}(2)_L$. We may identify the Goldstones $h^m$ with the usual elementary Higgs multiplet as
\begin{equation}
\label{eq:Hh}
 H=\left[
 \begin{array}{c}
 H^+\\
  H^0
 \end{array}
 \right]=\frac{1}{\sqrt{2}}\left[
 \begin{array}{c}
  h^1+ih^2\\
  h^3+ih^4
 \end{array}
 \right].
\end{equation}
In the Higgs vacuum we have respectively $\langle h\rangle=\langle h^3\rangle\neq0$, and 
\begin{align}
\label{eq:vacuum}
 \langle\Sigma^\top\rangle=[0,0,\epsilon,0,\sqrt{1-\epsilon^2}],\quad\epsilon=\sin\frac{\langle h\rangle}{f_\pi}.
\end{align}
The covariant derivatives in the Lagrangian of Eq.~\eqref{eq:L} include SU(2)$_L$ generators. From the identification of Eq.~\eqref{eq:Hh} we may construct the representation of SU(2)$_L$ on the Goldstone
multiplet $\Sigma$ as follows,
\begin{equation}
\label{eq:SU2gen}
 \begin{aligned}
 D_\mu\Sigma=&\partial_\mu\Sigma-i A_\mu^a T^a\Sigma,\\
 T^1=&\frac{i}{2}\left[
 \begin{array}{ccccc}
 0 & 0 & 0 & 1 & 0\\
 0 & 0 & -1 & 0 & 0\\
 0 & 1 & 0 & 0 & 0\\
 -1 & 0 & 0 & 0 & 0\\
  0 & 0 & 0 & 0 & 0
 \end{array} \right],
 \,T^2=&\frac{i}{2}\left[
 \begin{array}{ccccc}
 0 & 0 & -1 & 0 & 0\\
 0 & 0 & 0 & -1 & 0\\
 1 & 0 & 0 & 0 & 0\\
 0 & 1 & 0 & 0 & 0\\
  0 & 0 & 0 & 0 & 0
 \end{array} \right],
 \,T^3=&\frac{i}{2}\left[
 \begin{array}{ccccc}
 0 & 1 & 0 & 0 & 0\\
 -1 & 0 & 0 & 0 & 0\\
 0 & 0 & 0 & -1 & 0\\
 0 & 0 & 1 & 0 & 0\\
  0 & 0 & 0 & 0 & 0
 \end{array} \right].
 \end{aligned}
\end{equation}
Note how the generators are a subset of the unbroken SO(4) symmetry acting on the first four components of $\Sigma$.
The mass of the gauge bosons in the vacuum is defined by
\begin{equation}
 m^2_W=\frac{g^2\epsilon^2 f^2_\pi}{4}\equiv\frac{g^2v^2}{4},
 \label{eq:mw}
\end{equation}
where at tree-level $v= \epsilon f_\pi = 246$ GeV. Note that $v$ here does not represent the pseudo-Goldstone VEV, but rather parameterizes the $W$ mass. For $|\alpha/(2\beta)|\leq1$ the scalar potential has a minimum at
\begin{equation}
\label{eq:vev2}
 \cos\frac{\langle h\rangle}{f_\pi}=-\frac{\alpha}{2\beta},
\end{equation}
and the fluctuations of the field $h^3$ -- the  pseudo-Goldstone version of the Higgs -- acquire a mass
\begin{align}
 m^2_h=\frac{2\beta\epsilon^2}{f_\pi^2}.
\end{align}
Note that a positive Higgs mass requires $\beta>0$, while Eq.~\eqref{eq:vev2} will have a solution for either positive or negative $\alpha$. It appears that there are two families of solutions, but as we will
argue later they are physically equivalent. Beyond the known value of the weak gauge coupling, the bosonic low-energy Lagrangian of Eq.~\eqref{eq:L} has three parameters $f_\pi,\alpha,\beta$, and there is freedom in choosing the sign of $\alpha$. Requiring that the masses of the gauge bosons and the Higgs 
reproduce their measured values leaves one free parameter, which we take as $f_\pi$. Thus we may write $\alpha,\beta$ in terms of the physical masses $m_W,m_h$ and $f_\pi$:
\begin{align}
\label{eq:abeta}
 \alpha=\pm\frac{g f_\pi^4}{4}\frac{m^2_h}{m_W}\sqrt{\frac{g^2}{m^2_W}-\frac{4}{f_\pi^2}},\quad \beta=\frac{g^2 f^4_\pi}{8}\frac{m^2_h}{m^2_W}.
\end{align}
Note that consistency demands
\begin{align}
\label{eq:fmin}
 f_\pi > \frac{2m_W}{g}.
\end{align}
Picking the lowest possible value of $\langle h\rangle$ yielding the correct gauge boson mass, (i.e. $\sin h/f_\pi>0$) one has
\begin{equation}
\label{eq:vev}
\begin{aligned} 
 &\langle h\rangle_-=v+\frac{v^3}{6f_\pi^2}+\frac{3v^5}{40f_\pi^4}+{\cal O}\left(\frac{1}{f_\pi^6}\right)~~\mathrm{for}~~\alpha<0,\\
 &\langle h\rangle_+=\pi f_\pi-\langle h\rangle_-=\pi f_\pi-v-\frac{v^3}{6f_\pi^2}-\frac{3v^5}{40f_\pi^4}+{\cal O}\left(\frac{1}{f_\pi^6}\right)~~\mathrm{for}~~\alpha>0,
\end{aligned}
\end{equation}
which shows explicitly the modified relation between the Higgs VEV and the $W$ masses alluded to before. At this point, one would be tempted to argue that models in the large $f_\pi$, $\alpha>0$ branch, 
 with $\langle h\rangle \gg v$, should develop a larger sphaleron barrier. In previous sections we saw that the sphaleron configurations probe
the potential between the origin and the Higgs VEV, and so for large $\langle h\rangle$ the sphaleron profile would have to cover a larger amount of field-space,
implying larger kinetic contributions to the bosonic energy. Alas, this intuition is misleading. An important difference with respect to the elementary Higgs case is that the theory has a discrete selection rule.
It can be easily seen that in a unitary gauge with $h^i=0,i\neq3$, and thus $h^3=h$, the Lagrangian of Eq.~\eqref{eq:L} is invariant under the discrete transformation
\begin{align}
\label{eq:alphalaw}
\alpha\rightarrow-\alpha,\quad h\rightarrow \pi f_\pi-h.
\end{align}
This means that sphaleron solutions for one choice of sign of $\alpha$ can be related to sphaleron solutions for the other choice,  with matching energies. For this reason the two choices of sign of $\alpha$ are physically equivalent.  This equivalence can also be seen by  expanding the Lagrangian in the unitary gauge around the vacuum configurations, $ h=\langle h\rangle+\delta h$,  with $\langle h\rangle$ given in Eq.~\eqref{eq:vev}, and with $\alpha$ either positive or negative as in Eq.~\eqref{eq:abeta}. Doing  a $1/f_\pi$ expansion, the resulting terms are related by an unphysical $Z_2$ transformation $\delta h\rightarrow-\delta h$.  The same conclusion applies to fermionic couplings, which we did not discuss here but can be modelled again with $\sin h/f$ interactions \cite{Agashe:2004rs}. Therefore both scenarios with $\alpha>0$ and $\alpha<0$ are indistinguishable, and  we will focus on the $\alpha<0$ case.  We show in figure \ref{fig:Vcomp} the potential energy density of $h$ for $f=260$ GeV, showing the two equivalent realizations with different signs of $\alpha$. Note the different position of the VEVs and how the potentials are related by the transformation \eqref{eq:alphalaw}.

Given the row of zeros in the SU(2)$_L$ generators in Eq.~\eqref{eq:SU2gen}, it is clear that $\Sigma$ involves two irreducible representations. We may write $\Sigma=\tilde\Sigma\oplus\Lambda$, with
\begin{align}
\label{eq:tildeSigma}
 \tilde\Sigma^\top=\frac{\sin\frac{h}{f_\pi}}{h}[h^1,h^2,h^3,h^4],\quad
\Lambda=\cos\frac{h}{f_\pi}.
\end{align}
Then $D_\mu\tilde\Sigma=\partial_\mu\tilde\Sigma-iA^a_\mu\tilde T^a\tilde\Sigma$, with the $\tilde T^a$ given by the upper-left $4\times4$ blocks of the generators in Eq.~\eqref{eq:SU2gen}. In this way 
we can rewrite Eq.~\eqref{eq:L} as
\begin{equation}
\label{eq:L2}
  {\cal L}= \,\frac{f_\pi^2}{2}\partial_\mu\Lambda\partial^\mu\Lambda+\frac{f_\pi^2}{2}(D_\mu\tilde\Sigma)^\top D^\mu\tilde\Sigma-\frac{1}{4g^2}W^a_{\mu\nu}W^{a,\mu\nu}-\alpha\cos\frac{h}{f_\pi}+\beta\sin^2\frac{h}{f_\pi}.
\end{equation}
We are now ready to define the bosonic potential energy relevant for sphaleron configurations. As in the previous cases, this is just given by the Hamiltonian in the temporal $A_0=0$ gauge, with time derivatives
omitted (or equivalently, the Euclidean Lagrangian evaluated in static configurations). Performing rescalings as in Eq.~\eqref{eq:rescaling},  with $h_i\rightarrow m_W/g h_i$,  the dimensionless bosonic energy functional becomes
\begin{equation}
 \tilde V_{\rm bos}=\frac{1}{g^2}\int\!d^3y\,\left\{\frac{1}{4}\tilde W^a_{ij}\tilde W^{a}_{ij}+\frac{\hat f_\pi^2}{2}\partial_i\Lambda \partial_i\Lambda+\frac{\hat f_\pi^2}{2}(D_i\tilde\Sigma)^\top D_i\tilde\Sigma+\hat\alpha\cos\frac{\tilde h}{\hat f_\pi}-\hat\beta\sin^2\frac{\tilde h}{\hat f_\pi}\right\},
\end{equation}
where we defined the following modified couplings,
\begin{align}
 \hat f_\pi\equiv\frac{g f_\pi}{m_W},\quad \hat \alpha\equiv\frac{\alpha g^2}{m^4_W},\quad \hat \beta\equiv\frac{\beta g^2}{m^4_W}.
\end{align}
The equations for the sphaleron configurations that extremise $\tilde V_{\rm bos}$ are:
\begin{equation}
\label{eq:eom}
\begin{aligned}
({\cal D}_j W_{ij})^a+\frac{i\hat f_\pi^2}{2}\left(\tilde\Sigma^\top T^a D_i\tilde\Sigma-(D_i\tilde\Sigma)^\top T^a \Sigma\right)=0,\quad i,a=1,\dots,3,\\
\hat f^2_\pi\left(\partial^2\Lambda\frac{\partial\Lambda}{\partial h^m}+(D^2\tilde\Sigma)^n\frac{\partial\tilde\Sigma^n}{\partial h^m}\right)+\frac{h^m}{h\hat f_\pi}\sin\frac{h}{\hat f_\pi}
\left(\hat\alpha+2\hat\beta\cos\frac{h}{\hat f_\pi}\right)=0,\quad m=1,\dots,4,
\end{aligned} 
\end{equation}
with
\begin{align}
\frac{\partial\Lambda}{\partial h^m} &=-\frac{h^m}{h\hat f_\pi}\sin\frac{h}{\hat f_\pi},& \frac{\partial\tilde\Sigma^n}{\partial h^m} &=\frac{1}{h^3\hat f_\pi}\left(h \cos\frac{h}{\hfpi}
 -\hfpi\sin\frac{h}{\hfpi}\right)h^m h^n+\frac{\delta^{mn}}{h}\sin\frac{h}{\hfpi}.
\end{align}
We shall proceed as before and introduce the same rotationally symmetric ansatz as in Eq.~\eqref{eq:ansatz}, with $\tilde H$ interpreted in terms of the dimensionless Goldstone fields $\tilde h^i$ as
$\tilde H=[\tilde h_1+i\tilde h_2, \tilde h_3+i\tilde h_4]^\top$.
After introducing the ansatz in the equations of motion, and going as before into the $C=0$ gauge, one gets four independent equations of motion, which can be written as:
\begin{equation}
\label{eq:Escomp}
 \begin{aligned}
 &\,B''-\frac{B}{r^2}\left(A^2+B^2-1\right)+{\cal F}^2_1\left[2GF-B(G^2+F^2)\right]=0,\\
 &\,A''-\frac{A}{r^2}\left(A^2+B^2-1\right)-{\cal F}^2_1\left[A(G^2+F^2)+G^2-F^2\right]=0,\\
 \end{aligned}
 \end{equation}
 \begin{equation}
 \begin{aligned}
 &\,{{{\cal F}_1}}\left\{\hfpi^2 \left[{{\cal F}_2} G \left(4 F \left(2 B G+r \left(r F''+2 F'\right)\right)-G^2 \left(2 A^2+4 A+2 B^2+2\right)-F^2 \left(2 A^2-4 A+2 B^2+2\right)\right.\right.\right.\\
 &\left.\left.+4  r G\left(r G''+2 G'\right)\right)+{{\cal F}_1} F \left(F \left(4 r^2 G''+8 r G'-8 A G\right)-4 G^2 B+4F^2 B-4 G r \left(r F''+2 F'\right)\right)\right]\\
 &\left.+4 \hat\alpha G^3 r^2+4 \hat\alpha G F^2 r^2+8 \hat\beta {{\cal F}_2} G r^2 \left(G^2+F^2\right)\right\}\\
 &+{2\hfpi^2 \,r}\left\{{{\cal F}_1}' \left[4 {{\cal F}_2} G\left(G^2+F^2\right) +4 r G' \left({{\cal F}_1} F^2+{{\cal F}_2} G^2\right)+4 r G F  ({{\cal F}_2}-{{\cal F}_1}) F'\right]+2G \left(G^2+F^2\right)( r{{\cal F}_2}  {{\cal F}_1}'' \right.\\
 &\left.-r {{\cal F}_1} {{\cal F}_2}'' -2 {{\cal F}_1}  {{\cal F}_2}' )\right\}=0
,\\
\ \\
&\,{{{\cal F}_1}}\left\{\hfpi^2 \left[{{\cal F}_2} F \left(4 F \left(2 B G+r(r F''+2  F')\right)-G^2 \left(2 A^2+4 A+2 B^2+2\right)-F^2 \left(2 A^2-4 A+2 B^2+2\right)\right.\right.\right.\\
&\left.\left.+4 r G \left(r G''+2 G'\right)\right)+ {{\cal F}_1} G \left(G(4r^2 F''+8 r F'+8 A F)+4 B G^2-4 B F^2-4 r F(r G''+2G')\right)\right]\\
&\left.+4 \hat\alpha G^2 F r^2+4 \hat\alpha F^3 r^2+8 \hat\beta {{\cal F}_2} F r^2 \left(G^2+F^2\right)\right\}\\
&+{2 \hfpi^2\, r}\left[ {{\cal F}_1}' \left(4{{\cal F}_2} F\left(G^2+F^2\right) +4 r F' \left({{\cal F}_1} G^2+{{\cal F}_2} F^2\right)+4 G F r ({{\cal F}_2}-{{\cal F}_1}) G'\right)+2 F \left(G^2+F^2\right) \left(r{{\cal F}_2} {{\cal F}_1}''\right.\right.\\
&\left.\left.-r{{\cal F}_1}  {{\cal F}_2}''-2 {{\cal F}_1} {{\cal F}_2}'\right)\right]=0.
 \end{aligned}
\end{equation}
In the previous equations, we defined the ``form-factors''

\be\begin{aligned}
 {\cal F}_1\equiv&\,\frac{\hfpi\sin\frac{h}{\hfpi}}{h}=\frac{\hfpi}{2\sqrt{G^2+F^2}}\sin\left[\frac{2\sqrt{G^2+F^2}}{\hfpi}\right],\\
  {\cal F}_2\equiv&\,\cos\frac{h}{\hfpi}=\cos\left[\frac{2\sqrt{G^2+F^2}}{\hfpi}\right].
\end{aligned}
\ee
These form factors ${\cal F}_1,{\cal F}_2$ encode the nontrivial interactions of the composite Higgs. They tend to $1$ for large $\hfpi$, for which one recovers the limiting case of the SM Higgs. This  can be explicitly 
checked from the above sphaleron equations, or by realizing that in this limit the bosonic Lagrangian of Eq.~\eqref{eq:L} coincides with the SM one.

The variables introduced in Eq.~\eqref{eq:RS} allow for a substantial simplification for the equations, which become
\begin{equation}
\label{eq:eqsR}
 \begin{aligned}
 & R''+\frac{\hfpi^2}{4}  \sin ^2\left[\frac{2 S}{\hfpi}\right] \cos [2 \phi-\theta]+R \left(\frac{\hfpi^2}{8}  \cos \left[\frac{4 S}{\hfpi}\right]-\frac{\hfpi^2}{8}+\frac{1}{r^2}-\theta'^2\right)-\frac{R^3}{r^2}=0,\\
  &2 r \left(r S''\!+2 S'\right)+\frac{\hfpi}{4}  \sin \left[\frac{4 S}{\hfpi}\right] \left(2 R \cos [2 \phi-\theta]\!-\!2 r^2 \phi'^2\!-\!R^2\!-\!1\right)+\frac{r^2}{\hfpi} \sin \left[\frac{2 S}{\hfpi}\right] \left({\hat\alpha}+2 {\hat\beta} \cos \left[\frac{2 S}{\hfpi}\right]\right)=0,\\
 & 4 R \theta''+8 R' \theta'+\hfpi^2 \sin ^2\left[\frac{2 S}{\hfpi}\right] \sin [2 \phi-\theta]=0,\\
  & r \left(\hfpi r \phi''+2 \phi' \left(2 r S' \cot \left[\frac{2 S}{\hfpi}\right]+\hfpi\right)\right)-\hfpi R \sin [2 \phi-\theta]=0.
 \end{aligned}
\end{equation}
In these variables we may write $\tilde V_{\rm bos}$ as
\begin{equation}
\label{eq:Vbospol}
 \begin{aligned}
  \tilde V_{\rm bos}=&\frac{4\pi}{g^2}\int dr\left\{\frac{1}{4} \left[\hfpi^2 \sin ^2\left[\frac{2 S}{\hfpi}\right] \left((R-1)^2+2 R(1- \cos [2 \phi-\theta])\right)+4 R^2{ \theta'}^2+4 {R'}^2\right]\right.\\
  &\left.+\frac{1}{2} \left(R^2-1\right)^2+\frac
 {r^2}{4}\left[\frac{1}{\beta}\left(\hat\alpha+2 \hat\beta \cos \left[\frac{2 S}{\hfpi}\right]\right)^2+2 \hfpi^2 {\phi'}^2 \sin ^2\left[\frac{2 S}{\hfpi}\right]+8 {S'}^2\right]
  \right\}.
 \end{aligned}
\end{equation}
As in the previous cases, the last two equations in \eqref{eq:eqsR} can be solved as in  \eqref{eq:thetas}. This  gives the simplified system 
\begin{equation}
\label{eq:eqsRsimple}
 \begin{aligned}
 &R''+ R \left(\frac{1}{r^2}-\frac{1}{4} \hfpi^2 \sin ^2\left[\frac{2 S}{\hfpi}\right]\right)\pm\frac{1}{4} \hfpi^2 \sin ^2\left[\frac{2 S}{\hfpi}\right]-\frac{R^3}{r^2}=0,\\
 &r^2 S''+2 r S'+\frac{1}{8\hfpi}\left(4 \hat\alpha r^2 \sin \left[\frac{2 S}{\hfpi}\right]-\sin \left[\frac{4 S}{\hfpi}\right] \left(\hfpi^2 (R\mp1)^2-4 \hat\beta r^2\right)\right)=0.
 \end{aligned}
\end{equation}
The aforementioned physical equivalence between models with  $\alpha>0$ and $\alpha<0$ can be understood from the bosonic energy \eqref{eq:Vbospol} and equations \eqref{eq:eqsR}, \eqref{eq:eqsRsimple} by noting that they are invariant under the discrete symmetry
\begin{align}
\hat\alpha\rightarrow-\hat\alpha,\quad S\rightarrow S+\frac{\pi}{2} \hfpi.
\end{align}
Thus, solutions with one sign of $\hat\alpha$ can always be mapped onto solutions with the other sign. With the sign of $\hat\alpha$ fixed, another discrete symmetry of the bosonic energy and the sphaleron equations is
\begin{align}
\label{eq:discretesym}
 S\rightarrow \pi \hfpi\pm S.
\end{align}

 The asymptotic solutions for Eq.~\eqref{eq:Escomp} in the limit of  
large and small $r$ are given in appendix \ref{app:asympt}, and for fixed $f_\pi$ depend on the same number of parameters as in cases with an  elementary Higgs. The corresponding  solutions for \eqref{eq:eqsR} and \eqref{eq:eqsRsimple} can be obtained by using the definitions in \eqref{eq:RS}. Once again, reconstructing the full profile of the sphaleron from the solutions to the simplified system \eqref{eq:eqsRsimple} requires to fix the ambiguity in the solution \eqref{eq:thetas} for $\theta,\phi$. As in the SM, parity-invariant sphalerons are expected to have  $N_{CS}=1/2$, which fixes $\theta=\pi+n\pi$.  Fixing $R>0$ at large $r$, we have found solutions with $\theta=\pi$ in the upper branch of Eq.~\eqref{eq:eqsRsimple}, corresponding to $\phi=\pi/2$ (see \eqref{eq:thetas}).

A distinguishing feature of composite Higgs scenarios is that there are new types of asymptotic solutions at $r\rightarrow0$ that can support novel sphaleron solutions with $N_{CS}=1/2$. In the SM and in the case of a deformed potential, sphalerons ended up having $S(r=0)=0$, as can be seen in figures \ref{fig:profiles_SM} and \ref{fig:profiles_def}. The existence of new solutions with 
$S(0)\neq0$ can be understood as follows. By continuity with the SM case, one expects solutions with $S(0)=0$ for both $\hat\alpha>0$ and $\hat\alpha<0$. However, as was just argued, solutions with $\hat\alpha>0$  and $S(0)=0$ can be mapped to solutions with the opposite sign of $\hat\alpha$ by doing $S\rightarrow S+\pi/2 \hfpi$, so that one ends up with $S(0)=\pi/2\hfpi$. For $\hat\alpha<0$ this corresponds to a local maximum of the potential. 
We see that sphalerons interpolating between the minimum of the scalar potential and the origin, for a given choice of $\hat\alpha$, are equivalent to sphalerons that interpolate between the minimum and a local maximum for the opposite choice of $\hat\alpha$ (see figure \ref{fig:Vcomp}).  The existence of new solutions with different behaviour near $r=0$  implies that there must be a new family of asymptotic solutions for small $r$, 
which is given in appendix \ref{app:asympt}. Amusingly, as further discussed in the appendix, although the choices of opposite values $\hat\alpha$ are equivalent, regular solutions for a given sign of $\hat\alpha$ correspond to singular solutions with the opposite $\hat\alpha$, although the singularity is unphysical, as it can be removed with a gauge transformation. The existence of a new family of sphalerons with $N_{CS}=1/2$ and for the observed value of the Higgs mass is a novel effect which is not present in models with elementary Higgses. In that case, as in the SM, new branches of sphalerons typically have $N_{CS}\neq 1/2$ and only appear if the Higgs is much more massive than observed.

We have computed numerically the sphaleron energy in both families of sphalerons, using the iterative method described in previous sections. The solutions using the four differential equations \eqref{eq:eqsR} confirm the constant values of the angles, $\theta=\pi=2\phi$, derived from \eqref{eq:thetas} and the requirement for $N_{CS}=1/2$. Restricting the analysis to $\hat\alpha<0$, the family of solutions with the usual $S(0)=0$ behaviour gives a sphaleron barrier which, as expected, recovers the SM result in the limit of large $f_\pi$. The new family of $N_{CS}=1/2$ solutions 
has greater energies.

\begin{figure}[h!]
\begin{center}
\includegraphics[width=0.48\textwidth]{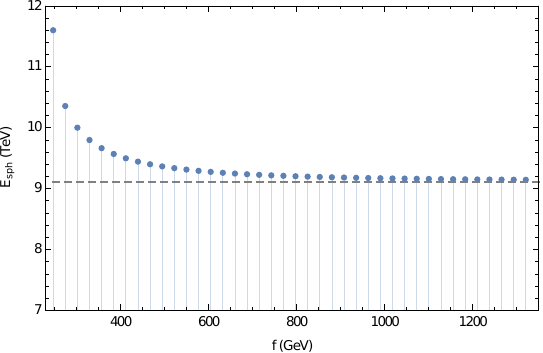}
\includegraphics[width=0.48\textwidth]{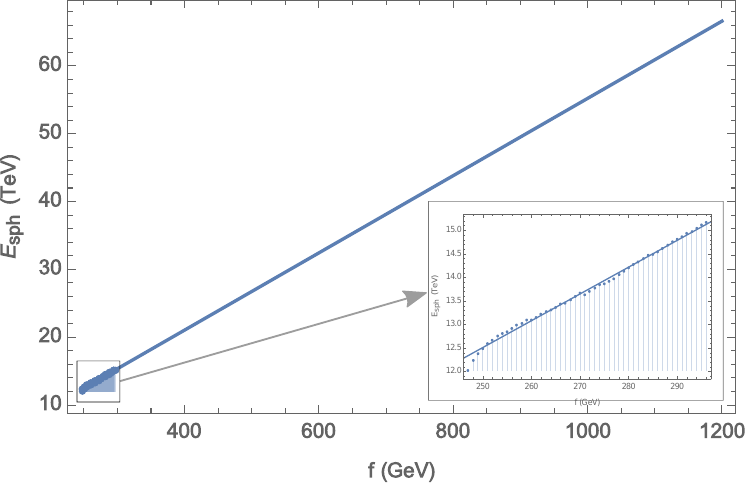}
\caption{Energy barrier between topological vacua as a function of $f_\pi$, in the $S(0)=0$  (left) and  $S(0)\neq0$ branches. The horizontal line marks the SM limit.}
\label{fig:barrier_smallh}       
\end{center}
\end{figure}

\begin{figure}[h!]
\begin{center}
\includegraphics[width=0.49\textwidth]{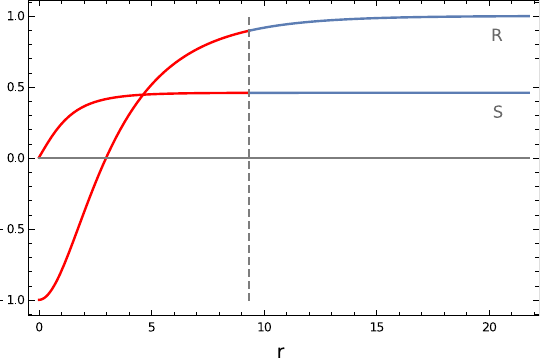}
\includegraphics[width=0.49\textwidth]{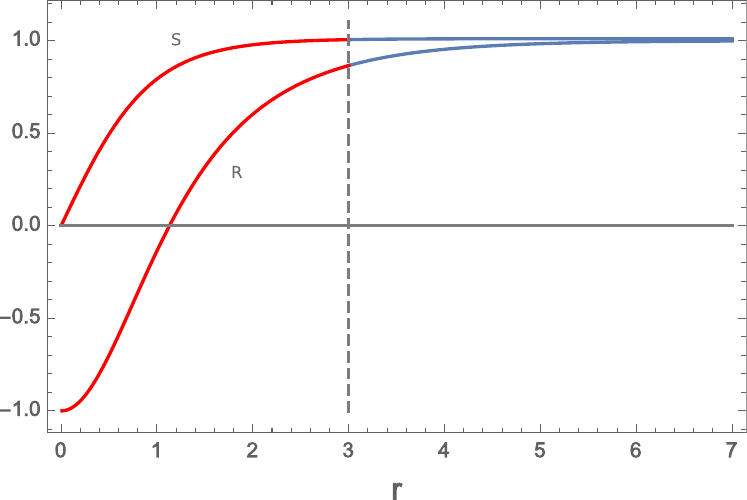}
\includegraphics[width=0.49\textwidth]{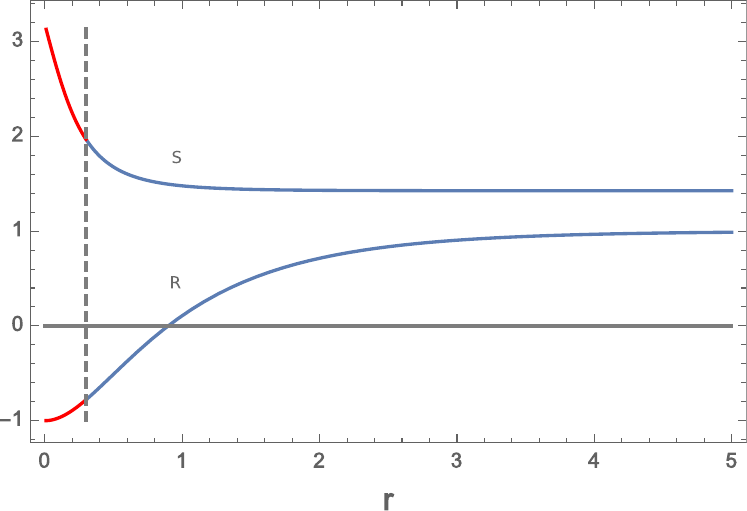}
\includegraphics[width=0.49\textwidth]{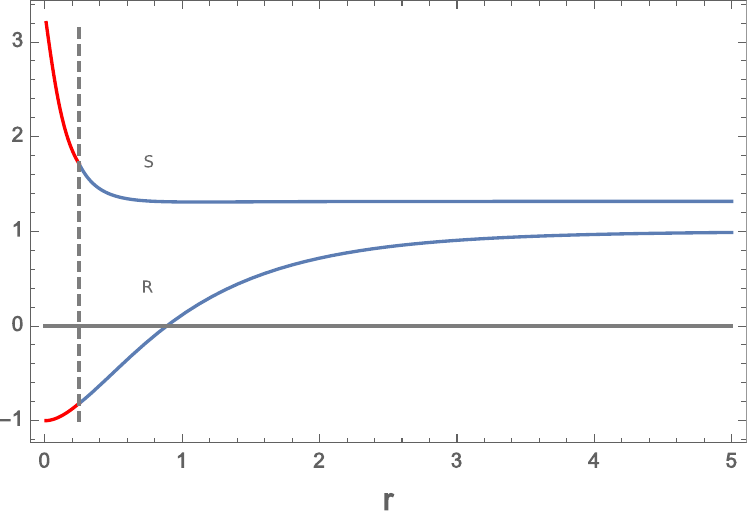}
\caption{Profile functions for $R,S$ in the sphaleron configurations obtained by solving the system of 2 differential equations. The vertical line marks the scale at which the low $r$ solution (red) was matched with the high $r$ solution (blue). Top: solutions in the $S(0)=0$ branch, with $f_\pi=250$ GeV (left) and $f_\pi=1$ TeV (right). Bottom: solutions in the $S(0)\neq0$ branch, with $f_\pi=250$ GeV (left) and $f_\pi=1$ TeV (right).}
\label{fig:profiles}       
\end{center}
\end{figure}

\begin{figure}[h!]
\begin{center}
\includegraphics[width=0.49\textwidth]{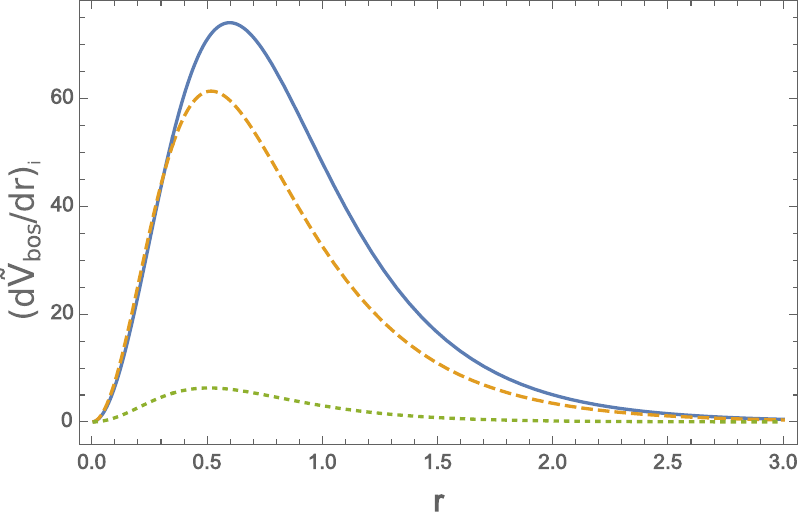}
\includegraphics[width=0.49\textwidth]{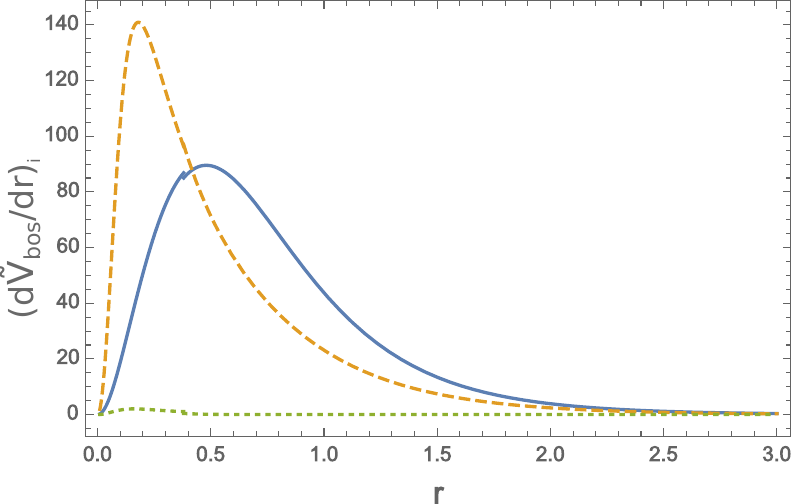}
\caption{Contributions to the integrand of the dimensionless bosonic energy functional for $\hfpi=260$ GeV in the $S(0)=0$  branch (left) and for $S(0)\neq0$ GeV (right). The contributions in solid blue are due to the gauge fields; those coming from derivatives of the scalar fields are shown with dashed orange lines, while those coming from the scalar potential 
energy density are shown with dotted green curves. }
\label{fig:Ecomp}       
\end{center}
\end{figure}

As anticipated earlier, the fact that in composite models not only the potential of the Higgs but also its derivative interactions are modified  allows for larger deviations from the value of $E_{\rm sph}$, even for the usual family with $S(0)=0$. In this case, $E_{\rm sph}$ reaches nearly 12 TeV for the theoretical  minimum 
$f_\pi=v$ (see equation \eqref{eq:fmin}), while it decreases rapidly with growing $f_\pi$. With current collider bounds demanding $f_\pi\gtrsim 0.5$ TeV \cite{Aad:2015pla, Azatov:2013hya}, the sphaleron barrier differs from the SM by less than three percent. In the $S(0)\neq0$ branch, the sphaleron barrier starts similarly at 12 TeV and grows linearly with $f_\pi$. The dependence of $E_{\rm sph}$   with the compositeness scale in the two branches is illustrated in Fig.~\ref{fig:barrier_smallh}. The energies in the $S(0)\neq0$ branch are subject to more numerical uncertainties due to the function $S$ becoming very steep at the origin, which prevents convergence of the iterative approach for large enough values of $f_\pi$. Still, our calculations show a linear growth which, when extrapolated, predicts a barrier of around 28 TeV for $f_\pi=500$ GeV and 70 TeV for $f_\pi=1.2$ TeV, respectively.
Example profiles for $R,S$ of the resulting solutions are shown in Fig.~\ref{fig:profiles}. Note the steepness of $S$ near the origin in the lower graphs corresponding to the  $S(0)\neq0$  branch, which affects numerical convergence. For the same examples, Fig.~\ref{fig:Ecomp} shows  the contributions to the integrand of the dimensionless bosonic energy functional  $\tilde V_{\rm bos}$.
For the $S(0)\neq0$  branch and for large enough $f_\pi$, $E_{\rm sph}$ becomes dominated 
by the scalar derivatives, in contrast to the cases with elementary fields (see figures \ref{fig:profiles_SM} and \ref{fig:profiles_def}). This is a consequence of the fact that, in this branch, the sphaleron profile interpolates between the electroweak vacuum and the maximum of the scalar potential at $\pi f_\pi$. The distance in field space travelled by the sphaleron increases linearly with  $f_\pi$,  and we observe the same for the integral yielding $E_{\rm sph}$.

\section{Summary}
\label{sec:summary}

In this paper we have investigated the size of the sphaleron energy barrier in non-standard realizations of the Higgs vacuum. The sphaleron energy, which in the Standard Model lies near 9 TeV, sets the scale of baryon number violating processes mediated by the B+L anomaly, and is sensitive to properties of the Higgs potential away from the electroweak minimum. Thus, sphaleron-induced processes -- which would manifest themselves through the production of a large number
of quarks and leptons -- could offer new perspectives on the nature of the Higgs particle, offering global information about its potential. This is in contrast to the perturbative processes usually considered at colliders, which only probe the Higgs interactions locally, i.e. at the electroweak vacuum. 

The sphaleron energy can also be affected by modifications of the derivative interactions of the Higgs. In order to quantify the possible variations of $E_{\rm sph}$ in non-standard scenarios, we have calculated
$E_{\rm sph}$ in models which exhibit either a modified potential for an elementary Higgs, or both a modified potential and modified derivative terms for a composite Higgs. Such examples capture quite generic possibilities for the Higgs interactions, and can be considered as benchmarks for the possible variations of the scale of baryon number violation processes in theories beyond the Standard Model. 

For an elementary Higgs with a modified potential, we considered a generic parametrization of the former involving a logarithmic term, which can  introduce a barrier with respect to the origin and 
modify the depth of the electroweak vacuum. For long-lived electroweak vacua we find deviations of the sphaleron barrier which are at most of the order of 10\%. Such small deviations become less surprising after realizing that already in the SM the sphaleron energy is dominated by derivative contributions, which mostly depend on the distance on field space covered by the scalar profile of the sphaleron. In models with an elementary Higgs, the sphaleron interpolates between the origin and the electroweak vacuum, whose position is fixed by the masses of the weak gauge bosons.

In composite Higgs models the situation could be in principle different, since both the derivative interactions and the relation between the weak boson masses and the Higgs VEV are modified.  We centered our study in minimal composite Higgs models, in which the Higgs is a pseudo-Goldstone of a global SO(5) symmetry broken down to SO(4).  We have found that, 
in contrast with models with elementary Higgses, for which there are no multiple sphaleron branches for the observed values of the Higgs and gauge boson masses, composite Higgs models exhibit at least two branches of sphaleron solutions. The existence of a new branch can be understood from the discrete translation symmetries of the effective action for the pseudo-Goldstone fields. In contrast to known non-standard sphaleron branches for heavy elementary Higgses, sphalerons in this new branch still have half-integer Chern-Simons number, and an energy higher than the sphalerons in the usual branch. In the latter, although large deviations of $E_{\rm sph}$ are possible at low values of the compositeness scale, they are ruled out by collider bounds, so that the minimum sphaleron energy can only differ from the SM one by less than 3\%. On the other hand, sphalerons in the new branch have an energy that grows linearly with the compositeness scale, and would reach around 28 TeV if extrapolated to $\hfpi=500$ GeV, and 70 TeV for $\hfpi=1.2$ TeV. The new branch of sphaleron configurations is suggestive of a new high-energy threshold for baryon-violating processes in addition to the SM-like threshold at 9 TeV.

Concerning the theoretical  precision of our calculations, it should be noted that we set the weak mixing angle $\theta_W$ to zero. In models with elementary Higgses, a nonzero $\theta_W$ is known to induce changes in the sphaleron energy of less than a percent \cite{Kleihaus:1991ks,Ahriche:2014jna}. Such modifications are smaller than the largest deviations of $E_{\rm sph}$ with respect to its SM value that were calculated in the models analyzed in this work. Hence, we expect our estimates to be robust with respect to the inclusion of mixing-angle effects.

\acknowledgements
We want to thank Valya Khoze and Kazuki Sakurai for very helpful discussions and comments. MS is supported in part by the European Commission through the ``HiggsTools'' Inital Training Network PITN-GA-2012-316704. 

\appendix
 
 \section{Asymptotic solutions for large and small $r$\label{app:asympt}} 
 
 
 \subsection{Standard Model}
 
 At large $r$, finiteness of $V_{\rm bos}$ evaluated on the sphaleron implies that gauge fields must approach a pure gauge configuration, while the scalar fields must tend to a minimum of their potential. Starting from the topological gauge with $A_i\rightarrow0,\,r\rightarrow\infty$, one has
\begin{align}
\label{eq:topinfty}
 A(r)\rightarrow 1, \quad B(r)\rightarrow0,\quad C(r)\rightarrow0 \quad \text{for } r\rightarrow\infty, \text{ topological gauge}.
\end{align}
This in turn implies that $F(r)$ and $G(r)$ should approach the value that minimizes the Higgs potential, which is
\begin{align}
\label{eq:FGinfty}
 F(r)\rightarrow1, G(r)\rightarrow0.
\end{align}
As a consequence of \eqref{eq:gaugetr}, the $C=0$ gauge can be obtained by fixing the derivative of the local transformation parameter $P'=-C/(2r)$, which determines $P$ up to a constant of integration. Then the boundary conditions in the $C=0$ gauge, obtained by transforming \eqref{eq:topinfty} and \eqref{eq:FGinfty} in accordance with \eqref{eq:gaugetr}, end up depending on an arbitrary constant:
 \begin{equation}
\label{eq:CzeroinftySM}
\begin{aligned}
A(r)\rightarrow&\, \cos\theta_\infty, & B(r)\rightarrow&\, \sin\theta_\infty,\\
F(r)\rightarrow &\, \cos\frac{\theta_\infty}{2}, & G(r)\rightarrow&\,  \sin\frac{\theta_\infty}{2},    \quad \text{for } r\rightarrow\infty, \text{ $C=0$ gauge}.
\end{aligned}
\end{equation}
The asymptotic solutions of equations \eqref{eq:Es}, \eqref{eq:E4E5SM} for large $r$ and with the boundary conditions \eqref{eq:CzeroinftySM} depend on three free paramameters 
$c_a,\delta,c_h$, and have the form:
\be\begin{aligned}
 A(r)\sim&(1+a(r))\cos\theta_\infty-b(r)\sin\theta_\infty&  B(r)\sim&(1+a(r))\sin\theta_\infty+b(r)\cos\theta_\infty\\
 F(r)\sim&\left(1+\frac{f(r)}{r}\right)\cos\frac{\theta_\infty}{2}-\frac{g(r)}{r}\sin\frac{\theta_\infty}{2} & G(r)\sim&\left(1+\frac{f(r)}{r}\right)\sin\frac{\theta_\infty}{2}+\frac{g(r)}{r}\cos\frac{\theta_\infty}{2},
\end{aligned}\ee
with 
\be
\begin{aligned}
 a(r)=&c_a \cos \delta e^{-r}+{\cal O}(r^{-1})&  b(r)=&c_a \sin \delta e^{-r}+{\cal O}(r^{-1}),\\
 f(r)=&c_h e^{-\kappa r}+{\cal O}(r^{-1})&  g(r)=&-c_a \frac{\sin \delta}{r}+{\cal O}(r^{-2}).
\end{aligned}
\ee
Given the ansatz \eqref{eq:ansatz}, it can be seen that the following expansion gives a regular behaviour at the origin, and is compatible with scalar functions with nonzero first derivatives at the origin,
(a common feature of sphaleron solutions):
\be
\label{eq:seriesfun}
\begin{aligned}
 A(r)=&1+A_2 r^2 +A_4 r^4+{\cal O}(r^4),    & B(r)=& B_1 r+B_3 r^3+{\cal O}(r^4),  \\
 {F}(r)=&{F}_0+{F}_2 r^2+{F}_4 r^4+{\cal O}(r^4), & G(r)=& G_1 r+G_3 r^3+{\cal O}(r^4).
\end{aligned}
\ee
A series expansion of the equations  \eqref{eq:Es}, \eqref{eq:E4E5SM} allows to show that one can choose 3 independent constants $A_2,{F}_0, G_1$, with the rest satisfying the 
relations
\be
\begin{aligned}
 A_4=&\frac{1}{10} \left(A_2 {F}_0^2+3 A_2^2+2 G_1^2\right),& B_1=&0,\\
 B_3=&-\frac{G_1 {F}_0}{3} , & {F}_4=&\frac{{F}_0 }{480} \left(12 A_2^2+4 G_1^2 \left(3 \kappa ^2+2\right)+\left(3 {F}_0^4-4 {F}_0^2+1\right) \kappa ^4\right),\\
 G_3=&\frac{ G_1}{60} \left(12 A_2+{F}_0^2 \left(3 \kappa ^2+2\right)-3 \kappa ^2\right).
\end{aligned}
\ee
As can be seen from equation \eqref{eq:gaugetr}, a gauge transformation with gauge parameter $\Pi=\pi/2$ sends $A$ to $-A$, so solutions with $A(0)=-1$ instead of 1 in \eqref{eq:seriesfun} are also admissible: the apparent
singularity is just a gauge artifact. However, when substituting the corresponding expansion in the sphaleron equations, one gets scalar profiles near the origin with zero first derivatives, which do not give rise
to sphaleron solutions.

 \subsection{Standard Model with a deformed potential}
 
For large $r$ in the $C=0$ gauge, the same reasoning as in the SM case yields asymptotic solutions  that take  the same form as in \eqref{eq:topinfty}. At small $r$, doing the 
same series expansion as in \eqref{eq:seriesfun}, yields again solutions parameterized by three constants $A_2,{F}_0, G_1$, with the rest being determined by the following relations:

\be\begin{aligned}
 A_4=&\frac{1}{10} \left(A_2 {F}_0^2+3 A_2^2+2 G_1^2\right),\\
 B_1=&0,\\
 B_3=&-\frac{G_1 {F}_0}{3} , \\
 {F}_4=&-\frac{{F}_0}{480 \gamma _0^4 \gamma _2^3} \left.\Bigg\{-\gamma _2^3 \gamma _0^4 \left.[12 A_2^2+4 G_1^2 \left(3 \kappa ^2+2\right)+\left(3 {F}_0^4-4 {F}_0^2+1\right)
 \kappa ^4\right]+8 \hat{\beta } 
 \gamma _2 \gamma _0^2 \left.\bigg[6 G_1^2 \left.\bigg(\gamma _2^2 \left.\bigg(4 \gamma \right.\right.\right.\right.\\
 &\left.\left.+2 \gamma _0^2 \log \frac{\gamma _0}{\gamma _2}+3\right)+\gamma _0^2 {F}_0^4-4 \gamma _0^2 \gamma _2 {F}_0^2\right)+\kappa ^2 \left.\bigg((3 \gamma +2) \gamma _2^2+\gamma _0^2 {F}_0^8-\gamma _0^2 
 \left(6 \gamma _2+1\right) {F}_0^6+\gamma _2 {F}_0^4  \left.\bigg(3 (4 \gamma\right.\right.\\
 &\left.\left.\left.\left.+3) 
 \gamma _2+
 \gamma _0^2 \left(6 \gamma _2 \log \frac{\gamma _0}{\gamma _2}
 +5\right)\right)-2 \gamma _2^2 {F}_0^2 \left(7 \gamma +2 \gamma _0^2 
 \log \frac{\gamma _0}{\gamma _2}+5\right)\right)\right]+16 \hat{\beta }^2 \left.\bigg[-(3 \gamma +2)^2 \gamma _2^3\right.\right.\\
 &+2 \gamma _0^4 {F}_0^{10}-\gamma _2^2 {F}_0^4 
 \left(3 (4 \gamma +3)^2 \gamma _2+12 \gamma _2 \gamma _0^4  \log^2\frac{\gamma _0}{\gamma _2}+\gamma _0^2 \left(30 \gamma +12 (4 \gamma +3) \gamma _2 
 \log \frac{\gamma _0}{\gamma _2}+20\right)\right)\\
 &-\gamma _0^2 \gamma _2 {F}_0^8 \left(8 \gamma+\gamma _0^2 \left(4 \log \left(\frac{\gamma _0}{\gamma _2}
 \right)+9\right)+6\right)+2 \gamma _0^2 \gamma _2 {F}_0^6 \left(3 \gamma +6 \gamma _2 \left(4 \gamma +2 \gamma _0^2 \log \frac{\gamma _0}{\gamma _2}
 +3\right)+2\right)\\
 &\left.\left.+4 (3 \gamma +2) \gamma _2^3 {F}_0^2 \left(4 \gamma+2 \gamma _0^2 \log \frac{\gamma _0}{\gamma _2}+3\right)\right.\bigg]\right\},\\
 G_3=&\frac{G_1}{{60 \gamma _0^2 \gamma _2}}\left\{\gamma _2 \gamma _0^2 \left[12 A_2+{F}_0^2 \left(3 \kappa ^2+2\right)-3 \kappa ^2\right]+12 \hat{\beta } 
 \left.\bigg[(3 \gamma +2) \gamma _2+\gamma _0^2 {F}_0^4\right.\right.\\
 &\left.\left. -\gamma _2 {F}_0^2\left.\bigg(4 \gamma+2 \gamma _0^2 \log \frac{\gamma _0}{\gamma _2}+3\right)\right]\right\}
\end{aligned}
\ee
As in the SM case, there are also asymptotic solutions with $A(0)=-1$, gauge equivalent to regular configurations, but they don't give rise to sphaleron solutions.

 \subsection{Composite Higgs}
 
At large $r$ in the topological gauge, the gauge fields must approach zero and satisfy \eqref{eq:topinfty}. On the other hand, the scalar functions $F$, $G$ should minimize the potential in \eqref{eq:L}, which implies
\begin{align}
\label{eq:topinfty2}
 G(r)\rightarrow0,\quad {F}(r)=\frac{\hfpi}{2}\arccos\left[-\frac{\alpha}{2\beta}\right], \quad \text{for } r\rightarrow\infty, \text{ topological gauge}.
\end{align}
After a gauge transformation to the $C=0$ gauge, reasoning as was done for the SM yields the following modified boundary conditions,
\begin{equation}
\label{eq:Czeroinfty}
\begin{aligned}
A(r)\rightarrow&\, \cos\theta_\infty, & B(r)\rightarrow&\, \sin\theta_\infty,\\
{F}(r)\rightarrow &\,  \frac{\hfpi}{2}\arccos\left[-\frac{\alpha}{2\beta}\right]\cos\frac{\theta_\infty}{2}, & G(r)\rightarrow&\,  \frac{\hfpi}{2}\arccos\left[-\frac{\alpha}{2\beta}\right]\sin\frac{\theta_\infty}{2},    \quad \text{for } r\rightarrow\infty, \text{ $C=0$ gauge}.
\end{aligned}
\end{equation}
Defining 
\begin{align}
\label{eq:Sinfty}                                                                                                                                                                          
S_\infty\equiv\frac{\hfpi}{2}\arccos\left[-\frac{\alpha}{2\beta}\right] = \frac{\langle h\rangle}{2},                                                                                                                                                                                        
\end{align}
and writing the asymptotic solutions as follows,
\begin{equation}
\label{eq:inftyexpansions}
 \begin{aligned}
  A(r)=&(1+a(r))\cos\theta_\infty-b(r)\sin\theta_\infty,& B(r)=&(1+a(r))\sin\theta_\infty+b(r)\cos\theta_\infty,\\
  F(r)=&\left(S_\infty+\frac{f(r)}{r}\right)\cos\frac{\theta_\infty}{2}-\frac{g(r)}{r}\cos\frac{\theta_\infty}{2},& G(r)=&\left(S_\infty+\frac{f(r)}{r}\right)\sin\frac{\theta_\infty}{2}+\frac{g(r)}{r}\cos\frac{\theta_\infty}{2},
 \end{aligned}
\end{equation}
then the bounded solutions for $a(r),b(r),h(r),g(r)$ depend as in the previous cases on three parameters $c_a,\delta,c_h$, and take the form
\begin{equation}
\label{eq:inftyexpansions2}
 \begin{aligned}
  a(r)=&\,c_a \cos(\delta)e^{- r}+{\cal O}(r^{-1}),&   b(r)=&\,c_a \sin(\delta)e^{- r}+{\cal O}(r^{-1}),\\
  f(r)=&\,c_h e^{-\kappa r}+{\cal O}(r^{-1}),& g(r)=&\,-\frac{c_a S_\infty}{r}\sin(\delta)  e^{- r}+{\cal O}(r^{-2}).
 \end{aligned}
\end{equation}
Regularity near the origin enforces again the analytic expansion of equation \eqref{eq:seriesfun}, setting the function $C$ to zero. Once again the expansions depend now on three independent coefficients, $A_2,{F}_0,G_1$, with the following relations for the lowest orders:
\begin{equation}
\label{eq:zeroexpansions}
 \begin{aligned}
  A_4=&\,\frac{1}{80{F}^2_0}\left[\hfpi^2 \left(A_2 {F}_0^2+2 G_1^2\right)-\hfpi^2 \cos \left[\frac{4 {F}_0}{\hfpi}\right] \left(A_2 {F}_0^2+2 G_1^2\right)\right.\\
  &+8 A_2 {F}_0^2 \left(3 A_2\right)\Big],\\
 B_1=&\,0, \quad B_3=-\frac{\hfpi^2 }{12 {F}_0}\,G_1 \sin ^2\left[\frac{2 {F}_0}{\hfpi}\right],\\
 {F}_2=&\,\frac{1}{24 \hfpi {F}_0^2}\left[-2 \alpha {F}_0^2 \sin \left[\frac{2 {F}_0}{\hfpi}\right]+\sin \left[\frac{4 {F}_0}{\hfpi}\right] \left(3 \hfpi^2 G_1^2-2 \beta {F}_0^2\right)-12 \hfpi G_1^2 {F}_0\right],\\
 G_3=&\,\frac{G_1}{120 \hfpi^2 {F}_0^3} \left[-12 \hfpi^2 {F}_0 \left(G_1^2 \left(\cos \left[\frac{4 {F}_0}{\hfpi}\right]+4\right)-2 A_2 {F}_0^2\right)+15 \hfpi^3 G_1^2 \sin \left[\frac{4 {F}_0}{\hfpi}\right]+8 {F}_0^3 \cos \left[\frac{2 {F}_0}{\hfpi}\right] \bigg(\alpha\right.\\
 &\left.\left.+2 \beta  \cos \left[\frac{2 {F}_0}{\hfpi}\right]\right)-10 \hfpi {F}_0^2 \sin \left[\frac{2 {F}_0}{\hfpi}\right] \left(\alpha +2 \beta  \cos \left[\frac{2 {F}_0}{\hfpi}\right]\right)+\hfpi^4 {F}_0^3 \sin ^2\left[\frac{2 {F}_0}{\hfpi}\right]\right].
 \end{aligned}
\end{equation}
The former coefficients have a regular limit when ${F}_0\rightarrow0$, which is the limit of the ordinary branch of sphalerons. A special feature of the composite Higgs case is that there are new asymptotic solutions
at small $r$ which are gauge equivalent to regular solutions. These new solutions have nonzero first derivatives for the scalar profiles, and give rise to a new branch of $N_{CS}=1/2$ solutions. Indeed, as mentioned earlier,  although regularity of the ansatz 
\eqref{eq:ansatz} at $r=0$ apparently imposes $A(r)\rightarrow1$ at $r\rightarrow0$, as chosen in \eqref{eq:seriesfun}, the fact that $A$ is equivalent to $-A$ under a gauge transformation (see \eqref{eq:gaugetr}, with $P=\pi/2$), implies that solutions with $A(0)=-1$ are also admissible, being gauge equivalent to regular solutions. In contrast to the solutions given by equations \eqref{eq:seriesfun}, \eqref{eq:zeroexpansions}, 
the new ones are as follows:
\be
\label{eq:seriesfun2}
\begin{aligned}
 A(r)=&-1+A_2 r^2 +A_4 r^4+{\cal O}(r^4),    & B(r)={\cal O}(r^4),  \\
 {F}(r)=&{F}_0+{F}_1 r +{F}_3 r^3+{\cal O}(r^4), & G(r)={\cal O}(r^4),
\end{aligned}
\ee
with two independent constants, $A_2,{F}_1$:
\be
\begin{aligned}
 A_4=&\,-\frac{1}{30} \left(9 A_2^2+6 {F}_1^2\right),\\
 {F}_0=&\frac{\hfpi\pi}{2},\\
 {F}_3=&\frac{1}{30 \hfpi^2}(3 \alpha  {F}_1-6 A_2 \hfpi^2 {F}_1-6 \beta  {F}_1-16 {F}_1^3).
\end{aligned}
\ee
As mentioned in the main text, solutions with a given $\alpha$ can be mapped to equivalent solutions with $\alpha'=-\alpha$ by the transformation $S\rightarrow S+\hfpi \pi/2$. This maps boundary conditions with $S(0)=0$ --corresponding to regular gauge fields-- to boundary conditions with $S(0)\neq0$ --corresponding to fields with a spurious singularity, as was just discussed. In particular, the  family of sphaleron 
solutions recovering the SM result for large $f_\pi$ can be realized in terms of both regular or singular sphalerons, depending on the choice of $\alpha$.


\bibliography{references}

\end{document}